\documentclass[journal]{IEEEtran}

\ifCLASSINFOpdf
\else
   \usepackage[dvips]{graphicx}
\fi
\usepackage{url}
\usepackage{graphicx}
\usepackage{amssymb}
\usepackage{amsfonts}
\usepackage{xcolor}
\usepackage{amsmath}
\usepackage{commath}
\usepackage{caption}
\usepackage{subcaption} 
\usepackage{mwe}
\usepackage{esvect}

\usepackage[nolist,nohyperlinks]{acronym}
\usepackage{amsmath}
\usepackage{breqn}
\usepackage{balance}
\usepackage{cite}
\usepackage{framed}
\usepackage{siunitx}
\usepackage{soul}
\usepackage{multirow}
\usepackage{etoolbox}
\usepackage{blindtext}
\usepackage{booktabs}
\usepackage{footmisc} 
\usepackage{algorithm}
\usepackage{algpseudocode}
\usepackage{gensymb}
\usepackage{authblk}
\usepackage{lipsum,cuted}
\usepackage{mathtools}
\usepackage{float}
\usepackage{psfrag}
\usepackage{array}
\usepackage{algpseudocode}

\usepackage{standalone}

\usepackage{url}
\usepackage[bookmarks=false]{hyperref}
\hypersetup{
  hidelinks,
  colorlinks=false,
  breaklinks=true,
  bookmarksopen=false}
\usepackage[noabbrev]{cleveref}

\bstctlcite{IEEEexample:BSTcontrol}

\begin{acronym}[MACHU]
  \acro{iot}[IoT]{Internet of Things}
  \acro{cr}[CR]{Cognitive Radio}
  \acro{ofdm}[OFDM]{orthogonal frequency-division multiplexing}
  \acro{ofdma}[OFDMA]{orthogonal frequency-division multiple access}
  \acro{scfdma}[SC-FDMA]{single carrier frequency division multiple access}
  \acro{rfic}[RFIC]{radio frequency integrated circuit}
  \acro{rf}[RF]{radio frequency}
  \acro{sdr}[SDR]{Software Defined Radio}
  \acro{sdn}[SDN]{Software Defined Networking}
  \acro{su}[SU]{Secondary User}
  \acro{ra}[RA]{Resource Allocation}
  \acro{qos}[QoS]{quality of service}
  \acro{usrp}[USRP]{Universal Software Radio Peripheral}
  \acro{mno}[MNO]{Mobile Network Operator}
  \acro{mnos}[MNOs]{Mobile Network Operators}
  \acro{gsm}[GSM]{Global System for Mobile communications}
  \acro{tdma}[TDMA]{Time-Division Multiple Access}
  \acro{fdma}[FDMA]{Frequency-Division Multiple Access}
  \acro{gprs}[GPRS]{General Packet Radio Service}
  \acro{msc}[MSC]{Mobile Switching Centre}
  \acro{bsc}[BSC]{Base Station Controller}
  \acro{umts}[UMTS]{universal mobile telecommunications system}
  \acro{Wcdma}[WCDMA]{Wide-band code division multiple access}
  \acro{wcdma}[WCDMA]{wide-band code division multiple access}
  \acro{cdma}[CDMA]{code division multiple access}
  \acro{lte}[LTE]{Long Term Evolution}
  \acro{papr}[PAPR]{peak-to-average power rating}
  \acro{hn}[HetNet]{heterogeneous networks}
  \acro{phy}[PHY]{physical layer}
  \acro{mac}[MAC]{medium access control}
  \acro{amc}[AMC]{adaptive modulation and coding}
  \acro{mimo}[MIMO]{multiple-input-multiple-output}
  \acro{rats}[RATs]{radio access technologies}
  \acro{vni}[VNI]{visual networking index}
  \acro{rbs}[RB]{resource blocks}
  \acro{rb}[RB]{resource block}
  \acro{ue}[UE]{user equipment}
  \acro{cqi}[CQI]{Channel Quality Indicator}
  \acro{hd}[HD]{half-duplex}
  \acro{fd}[FD]{full-duplex}
  \acro{sic}[SIC]{self-interference cancellation}
  \acro{si}[SI]{self-interference}
  \acro{bs}[BS]{base station}
  \acro{fbmc}[FBMC]{Filter Bank Multi-Carrier}
  \acro{ufmc}[UFMC]{Universal Filtered Multi-Carrier}
  \acro{scm}[SCM]{Single Carrier Modulation}
  \acro{isi}[ISI]{inter-symbol interference}
  \acro{ftn}[FTN]{Faster-Than-Nyquist}
  \acro{m2m}[M2M]{machine-to-machine}
  \acro{mtc}[MTC]{machine type communication}
  \acro{mmw}[mmWave]{millimeter wave}
  \acro{bf}[BF]{beamforming}
  \acro{los}[LOS]{line-of-sight}
  \acro{nlos}[NLOS]{non line-of-sight}
  \acro{capex}[CAPEX]{capital expenditure}
  \acro{opex}[OPEX]{operational expenditure}
  \acro{ict}[ICT]{information and communications technology}
  \acro{sp}[SP]{service providers}
  \acro{inp}[InP]{infrastructure providers}
  \acro{mvnp}[MVNP]{mobile virtual network provider}
  \acro{mvno}[MVNO]{mobile virtual network operator}
  \acro{nfv}[NFV]{network function virtualization}
  \acro{vnfs}[VNF]{virtual network functions}
  \acro{cran}[C-RAN]{Cloud Radio Access Network}
  \acro{vran}[V-RAN]{Virtual Radio Access Network}
  \acro{bbu}[BBU]{baseband unit}
  \acro{bbus}[BBU]{baseband units}
  \acro{rrh}[RRH]{remote radio head}
  \acro{rrhs}[RRH]{Remote radio heads} 
  \acro{sfv}[SFV]{sensor function virtualization}
  \acro{wsn}[WSN]{wireless sensor networks} 
  \acro{bio}[BIO]{Bristol is open}
  \acro{vitro}[VITRO]{Virtualized dIstributed plaTfoRms of smart Objects}
  \acro{os}[OS]{operating system}
  \acro{www}[WWW]{world wide web}
  \acro{iotvn}[IoT-VN]{IoT virtual network}
  \acro{mems}[MEMS]{micro electro mechanical system}
  \acro{mec}[MEC]{Mobile edge computing}
  \acro{coap}[CoAP]{Constrained Application Protocol}
  \acro{vsn}[VSN]{Virtual sensor network}
  \acro{rest}[REST]{REpresentational State Transfer}
  \acro{aoi}[AoI]{Age of Information}
  \acro{lora}[LoRa\texttrademark]{Long Range}
  \acro{iot}[IoT]{Internet of Things}
  \acro{snr}[SNR]{Signal-to-Noise Ratio}
  \acro{cps}[CPS]{Cyber-Physical System}
  \acro{uav}[UAV]{Unmanned Aerial Vehicle}
  \acro{rfid}[RFID]{Radio-frequency identification}
  \acro{lpwan}[LPWAN]{Low-Power Wide-Area Network}
  \acro{lgfs}[LGFS]{Last Generated First Served}
  \acro{wsn}[WSN]{wireless sensor network} 
  \acro{lmmse}[LMMSE]{Linear Minimum Mean Square Error}
  \acro{rl}[RL]{Reinforcement Learning}
  \acro{nb-iot}[NB-IoT]{Narrowband IoT}
  \acro{lorawan}[LoRaWAN]{Long Range Wide Area Network}
  \acro{mdp}[MDP]{Markov Decision Process}
  \acro{ann}[ANN]{Artificial Neural Network}
  \acro{dqn}[DQN]{Deep Q-Network}
  \acro{mse}[MSE]{Mean Square Error}
  \acro{ml}[ML]{Machine Learning}
  \acro{cpu}[CPU]{Central Processing Unit}
  \acro{gpu}[GPU]{Graphics Processing Unit}
  \acro{tpu}[TPU]{Tensor Processing Unit}
  \acro{ddpg}[DDPG]{Deep Deterministic Policy Gradient}
  \acro{ai}[AI]{Artificial Intelligence}
  \acro{gp}[GP]{Gaussian Processes}
  \acro{drl}[DRL]{Deep Reinforcement Learning}
  \acro{mmse}[MMSE]{Minimum Mean Square Error}
  \acro{fnn}[FNN]{Feedforward Neural Network}
  \acro{eh}[EH]{Energy Harvesting}
  \acro{wpt}[WPT]{Wireless Power Transfer}
  \acro{dl}[DL]{Deep Learning}
  \acro{yolo}[YOLO]{You Only Look Once}
  \acro{mec}[MEC]{Mobile Edge Computing}
  \acro{marl}[MARL]{Multi-Agent Reinforcement Learning}
  \acro{aiot}[AIoT]{Artificial Intelligence of Things}
  \acro{ru}[RU]{Radio Unit}
  \acro{du}[DU]{Distributed Unit}
  \acro{cu}[CU]{Central Unit}
  \acro{api}[API]{Application Programming Interface}
  \acro{cf}[CF]{Carbon Footprint}
  \acro{ci}[CI]{Carbon Intensity}
  \acro{oran}[O-RAN]{Open Radio Access Network}
  \acro{flops}[FLOPs]{floating-point operations per cycle per core}
  \acro{FLOPS}[FLOPs]{floating-point operations}
  \acro{5g}[5G]{fifth-generation}
  \acro{embb}[eMBB]{enhanced Mobile Broadband}
  \acro{urllc}[URLLC]{Ultra Reliable Low Latency Communications}
  \acro{mmtc}[mMTC]{massive Machine Type Communications}
  \acro{hdd}[HDD]{hard disk drive}
  \acro{ssd}[SSD]{solid state drive}
  \acro{fspl}[FSPL]{free space path loss}
  \acro{cav}[CAV]{Connected Autonomous Vehicle}
  \acro{xr}[XR]{Extended Reality}
  \acro{ap}[AP]{access point}
  \acro{ble}[BLE]{Bluetooth low energy}
  \acro{uwb}[UWB]{ultra wide band}
  \acro{rss}[RSS]{received signal strength}
  \acro{nn}[NN]{Neural Network}
  \acro{v2x}[V2X]{vehicle-to-everything}
  \acro{gai}[GAI]{Generative AI}
  \acro{aiaas}[AIaaS]{AI as a service}
  \acro{mlp}[MLP]{Multilayer Perceptron}
  \acro{ble}[BLE]{Bluetooth Low Energy}
  \acro{lorawan}[LoRaWAN]{Long Range Wide Area Network}
  \acro{FPGAs}[FPGAs]{Field Programmable Gate Arrays}
  \acro{llm}[LLM]{large language model}
  \acro{fl}[FL]{federated learning}
  \acro{pue}[PUE]{Power Usage Effectiveness}
  \acro{cue}[CUE]{Carbon Usage Effectiveness}
  \acro{wue}[WUE]{Water Usage Effectiveness}
  \acro{ee}[EE]{energy efficiency}
  \acro{kpi}[KPI]{Key Performance Index}
  \acro{dp}[DP]{Data Plane}
  \acro{cp}[CP]{Control Plane}
  \acro{aes}[AES]{Advanced Encryption Standard}
  \acro{osi}[OSI]{Open Systems Interconnection}
  \acro{cnn}[CNN]{Convolutional Neural Network}
  \acro{kan}[KAN]{Kolmogorov–Arnold Network}
  \acro{gadf}[GADF]{Gramian Angular Difference Field}
  \acro{PUE}[PUE]{Power Usage Effectiveness}
  \acro{CUE}[CUE]{Carbon Usage Effectiveness}
  \acro{WUE}[WUE]{Water Usage Effectiveness}
  \acro{aes}[AES]{Advanced Encryption Standard}
  \acro{ihrrc}[IHRLLC]{immersive, hyper reliable, and low-latency communication}
  \acro{phy}[PHY]{physical layer}
  \acro{doa}[DoA]{direction of arrival}
  \acro{mmwave}[mmWave]{millimeter-wave} 
  \acro{thz}[THz]{terahertz}
  \acro{csi}[CSI]{channel state information}
  \acro{music}[MUSIC]{Multiple Signal Classification}
  \acro{esprit}[ESPRIT]{Estimation of Signal Parameters via Rotational Invariance Techniques}
  \acro{mvdr}[MVDR]{Minimum Variance Distortionless Response}
  \acro{aoa}[AoA]{Angle-of-arrival}
  \acro{fspl}[FSPL]{free-space path loss}
  \acro{sr}[SR]{Symbolic Regression}
  \acro{ris}[RIS]{reconfigurable intelligent surface}
  \acro{gp}[GP]{Genetic Programming}
  \acro{crlb}[CRLB]{Cramér-Rao Lower Bound}
  \acro{fim}[FIM]{Fisher–information matrix}
  \acro{dnn}[DNN]{deep neural network}
  \acro{dl}[DL]{Deep Learning}
  \acro{miso}[MISO]{Multi-input-single-output}
  \acro{lstm}[LSTM]{long-short term memory}
  \acro{3gpp}[3GPP]{3rd generation partnership project}
  \acro{mae}[MAE]{Mean Absolute Error}
  \acro{rmse}[RMSE]{root mean square error}
  \acro{cdf}[CDF]{Cumulative Distribution Function}
  \acro{dsr}[DSR]{Deep Symbolic Regression}
  \acro{rnn}[RNN]{Recurrent Neural Network}
  \acro{rspg}[RSPG]{Risk-Seeking Policy Gradient}
  \acro{pqt}[PQT]{Priority Queue Training}
  \acro{nlos}[NLOS]{Non-Line-of-Sight}
\end{acronym}

\usepackage{orcidlink}

\begin{document}
\title{SABER: Symbolic Regression-based Angle of Arrival and Beam Pattern Estimator}

\author{
	\vskip 1em
	
	Shih-Kai~Chou$^{\orcidlink{0000-0002-9985-7340}}$,~\IEEEmembership{Member,~IEEE,} Mengran~Zhao$^{\orcidlink{0000-0002-8834-8639}}$,~\IEEEmembership{Member,~IEEE,} Cheng-Nan~Hu$^{\orcidlink{0000-0002-8753-3576}}$,~\IEEEmembership{Senior Member,~IEEE,} Kuang-Chung~Chou,~\IEEEmembership{Member,~IEEE,} Carolina~Fortuna$^{\orcidlink{0000-0003-0547-3520}}$, and Jernej~Hribar$^{\orcidlink{0000-0002-9446-7917}}$,~\IEEEmembership{Member,~IEEE}
	
	\thanks{
		Manuscript received xx xxxxx 2025; revised xx xxxxx 2025; accepted xx xxxxx 2025.
	This work was supported in part by the European Commission NANCY project (No.101096456) and by the Slovenian Research Agency under grants P2-0016 and J2-50071. \emph{(Corresponding author: Shih-Kai Chou)}.
		
		S.-K. Chou,  C. Fortuna and J. Hribar are with Jo\v{z}ef Stefan Institute, Ljubljana, Slovenia. (e-mail:\{shih-kai.chou, carolina.fortuna, jernej.hribar\}@ijs.si). 
		
		M. Zhao is with the Centre for Wireless Innovation, Queen’s University Belfast, BT3 9DT Belfast, U.K. He is also with the School of Information and Communication Engineering, Xi'an Jiaotong University, 710049 Xi'an, China. (e-mail: mengran.zhao@qub.ac.uk).
		
		C.-N. Hu and K.-C. Chou are with Asia Eastern University of Science and Technology, New Taipei City, Taiwan. (e-mail: fo011@mail.aeust.edu.tw, kcchou24g@gmail.com).
	}
}

\maketitle

\begin{abstract}
Accurate Angle-of-arrival (AoA) estimation is essential for next-generation wireless communication systems to enable reliable beamforming, high-precision localization, and integrated sensing. Unfortunately, classical high-resolution techniques require multi-element arrays and extensive snapshot collection, while generic Machine Learning (ML) approaches often yield black-box models that lack physical interpretability. To address these limitations, we propose a Symbolic Regression (SR)-based ML framework. Namely, Symbolic Regression-based Angle of
Arrival and Beam Pattern Estimator (SABER), a constrained symbolic-regression framework that automatically discovers closed-form beam pattern and AoA models from path loss measurements with interpretability. SABER achieves high accuracy while bridging the gap between opaque ML methods and interpretable physics-driven estimators. First, we validate our approach in a controlled free-space anechoic chamber, showing that both direct inversion of the known $\cos^n$ beam and a low-order polynomial surrogate achieve sub-0.5 degree Mean Absolute Error (MAE). A purely unconstrained SR method can further reduce the error of the predicted angles, but produces complex formulas that lack physical insight. Then, we implement the same SR-learned inversions in a real-world, Reconfigurable Intelligent Surface (RIS)-aided indoor testbed. SABER and unconstrained SR models accurately recover the true AoA with near-zero error. Finally, we benchmark SABER against the Cramér-Rao Lower Bounds (CRLBs). Our results demonstrate that  SABER is an 
interpretable and accurate alternative to state-of-the-art and black-box ML-based methods for AoA estimation.
\end{abstract}


\section{Introduction}
\label{sec:intro}

As the transition from 5G to 6G accelerates, the foundational pillars of 5G, which include \ac{embb}, \ac{urllc}, and \ac{mmtc}, will be significantly expanded and enhanced to support an even broader range of services and applications~\cite{URLLC_ref}. In particular, \ac{embb} will offer higher capacity and coverage; \ac{urllc} will evolve into \ac{ihrrc}, providing deterministic, resilient connectivity for time-critical domains such as self-driving vehicles and autonomous systems~\cite{shamsabadi2025exploring6gpotentialsimmersive}; and \ac{mmtc} will scale to connect vast numbers of devices in smart cities, digital twins, and pervasive sensing environments~\cite{di2024aipoweredurbantransportationdigital}.

To realize the expanded capabilities of \ac{embb}, \ac{ihrrc}, and \ac{mmtc} in 6G, directional communication is required. As 6G introduces new physical layer technologies that offer fine-grained control over spatial resources and operate at higher frequencies with tighter beamforming~\cite{mmwave_1}, alongside advanced techniques such as ultra (cell-free) massive \ac{mimo}~\cite{ummimo,ngo2024ultradensecellfreemassivemimo}, and \ac{ris}~\cite{RIS_ISAC}. Consequently, precise beam alignment and low-latency \ac{aoa} estimation are first-order requirements for robust links, tracking, handover, and spatial reuse. However, current \ac{aoa} estimators either depend on multi-antenna arrays and numerous snapshots~\cite{aoa_new}, which makes them 
sensitive to calibration and \ac{snr} and computationally expensive~\cite{aoa_expensive}. Alternatively, they rely on data-hungry and black-box \ac{ml} models with limited interpretability and cross-scenario generalization~\cite{ML_bbox}. Collectively, array/snapshot dependencies and black-box \ac{ml} models hinder reliable and interpretable \ac{aoa} estimation for timely beam alignment, tracking, and handover across deployment conditions. To overcome these limitations, we propose a physics-guided \ac{sr} framework that inverts beam patterns directly from the measured path loss coefficient, and further enables interpretable, \ac{ml}-based \ac{aoa} estimation.





\subsection{Classical AoA Estimating Techniques}
\label{sec:rw_Classic}
Traditionally, classical signal processing techniques such as \ac{music}~\cite{music_ref} and \ac{esprit}~\cite{esprit_ref} are used for \ac{aoa} estimation, which have been widely used due to their high resolution. These subspace-based algorithms decompose the covariance matrix of the received signals to identify the signal directions and offer precise \ac{aoa} estimates. However, they often require a large number of snapshots of Fourier measurements acquired from a uniform array of radiating elements and result in high computational complexity~\cite{downside_music_esprit}, making them difficult to apply in real time. In addition, beamforming techniques such as the \ac{mvdr} beamformer~\cite{MVDR_ref} aim to enhance signal reception from the desired directions while suppressing interference and noise, but they too reach their limits in dynamic and complex environments. To address these challenges, \ac{ml}-based techniques have been introduced to learn data-driven mappings for \ac{aoa} estimation. Building upon this paradigm, our work extends such approaches toward interpretable, physics-consistent modeling.

\subsection{ML-based Techniques}
\label{sec:rw_ML}

\ac{ai}/\ac{ml} approaches address the limitations of conventional beamforming by learning complex patterns from the measurement data. For instance, in \ac{mmwave} and sub-\ac{thz} scenarios,~\cite{AIinBP_1} proposed a \ac{ml}-based approach using only phaseless measurements of the received power, demonstrating a significant reduction in beam alignment overhead compared to exhaustive searches and an improvement over compressive sensing techniques in multipath environments~\cite{AI_BP1comp}. In addition,~\cite{AIinBP_3} investigated \ac{dl}-based techniques to predict transmitting beamforming based solely on historical \ac{csi} without requiring current channel information in a \ac{miso} downlink system. A \ac{lstm}-based channel prediction module was proposed to improve the performance of prediction. Furthermore, a comprehensive evaluation of \ac{ml}-based spatial-domain beam prediction and Time-Domain Beam prediction was carried out within a realistic \ac{3gpp}-compliant simulator~\cite{AIinBP_4}. It showed significant benefits in system-level \ac{kpi}, such as user throughput and measurement overhead. Moreover,~\cite{AIinBP_2} jointly optimizes probing-beam selection and beam prediction with an integrated neural network. Unlike traditional methods that utilize predefined and spatially regular beam patterns aimed to minimize the overhead and latency in a \ac{mmwave} \ac{mimo} downlink system. Their model consists of two jointly trained components: a sampling network that learns optimal sampling patterns through a differentiable approximation of the sampling function, and a beam prediction network. Inspired by the physics of antenna arrays, the prediction network uses \ac{cnn} and self-attention to select the best beams from the measured Reference Signal Received Power (RSRP). This adaptive method significantly outperforms conventional \ac{dnn} approaches that rely on static beam sampling configurations. Unlike the approach proposed in~\cite{AIinBP_2}, our model bypasses differentiable sampling and CNN/attention predictors, using symbolic regression to derive closed‑form, physics‑consistent inversions from a path‑loss coefficient. This yields an interpretable \ac{aoa} estimation.

\ac{ml}-based methods have also shown promise in \ac{aoa} estimation, reducing the need for extensive snapshots and computational resources. Particularly,~\cite{AI_AoA_review} highlighted that traditional \ac{aoa} estimation algorithms face several limitations, such as the need for prior knowledge of the number of signal sources, sensitivity to coherent signals, and performance degradation in noisy environments, while \ac{dl}-based methods address these challenges by reformulating \ac{aoa} estimation as a pattern recognition problem using neural networks to learn the mapping from signal data to arrival angles. Despite these methods are attractive, the performance is significantly affected by the \ac{snr}, the number of snapshots, antennas, and the number of signal sources. A hybrid approach was proposed in~\cite{AI_aoa_1} that combines the traditional \ac{aoa} estimation method, i.e., \ac{music}, with regression-based \ac{ml} models, such as Gaussian process and regression tree, to provide a more accurate \ac{aoa} estimation. Particularly, the framework first utilized the \ac{music} spatial spectrum as a feature extraction step, where the resulting spectrum data is then fed into a regression model. This hybrid \ac{music}-\ac{ml} framework can compensate the multi-path effects, due to the \ac{music} alone would provide inaccurate weighted average of all paths, in addition, using \ac{music}-processed data, the models need less input nodes, thereby reducing the computing complexity comparing using raw measurement data as input. The performance degradation caused by low \ac{snr} and an unknown number of sources has been studied in~\cite{AI_aoa_2,AI_aoa_5}. More specifically, authors in~\cite{AI_aoa_2} introduced a deep \ac{cnn} specifically designed for \ac{aoa} estimation in extremely low \ac{snr} environments. This is achieved by a \ac{cnn} with 2D convolutional layers trained on multi-channel data, which includes I/Q samples and phases. The results showed a more robust prediction at both low and high \ac{snr}s. However, while such deep learning models effectively learn statistical mappings from signal features to recover \ac{aoa}, they remain black-box estimators—offering limited interpretability and little insight into the underlying physical relationships between path loss, gain, and angle. In contrast, SABER leverages \ac{sr} to recover these dependencies in closed-form, providing both physical transparency and comparable accuracy.

Several new technologies also leverage \ac{ml}-based methods to achieve accurate \ac{aoa} estimation~\cite{ai_aoa_3, AI_aoa_4,AI_aoa_6,AI_others_AoA1,AI_others_AoA2}. For example, in \ac{ris}-aided systems,~\cite{AI_aoa_4} proposed directly embedded the \ac{ris} functionality into a neural network-based architecture, and create a learnable \ac{ris} layer, allowing a more efficient and robust control mechanism over classic maximum-likelihood approaches. \ac{dl} frameworks have proven particularly effective at learning complex mappings from signal data to arrival angles, with \ac{cnn} consistently outperforming classical spectral-based algorithms, especially in challenging low \ac{snr} environments and multi-path scenarios. However, these black-box approaches, while highly effective, offer limited insight into the underlying physical relationships governing the estimation process, making it difficult to understand, validate, or generalize the learned models beyond their training conditions. These limitations highlight the broader potential for \ac{ml} approaches that can discover hidden patterns and relationships in complex signal processing problems while maintaining interpretability, motivating the exploration of methods that can provide both accuracy and physical insight, such as \ac{sr}. Classical beam pattern and AoA estimation rely on explicit algorithms but suffer from high complexity, whereas \ac{ml} methods trade interpretability for flexibility. \ac{sr} bridges this divide by discovering closed-form, physics-consistent relations directly from data, offering a principled balance between analytical transparency and learning-based adaptability

\subsection{Contributions}
\label{sec:con}

In this work, we address the aforementioned gap by introducing a \ac{sr}-based framework for \ac{aoa} estimation. Namely, Symbolic Regression-based Angle of
Arrival and Beam Pattern Estimator (SABER), that operates directly on a single scalar feature: the measured path loss coefficient. which allows us to recover closed-form, physically interpretable beam-pattern expressions and \ac{aoa} estimations. 

Unlike conventional regression methods that assume a predefined equation, \ac{sr} uses evolutionary algorithms, such as \ac{gp}~\cite{SR_Generic}, to explore a vast space of possible functions, automatically deriving symbolic representations without prior knowledge of the underlying form. This approach is particularly important in the physics community, where complex, nonlinear relationships between variables challenge analytical modeling~\cite{anaqreh2024towards}. 
By leveraging \ac{sr}, we can efficiently approximate beam patterns and \ac{aoa} estimations, capturing crucial directional characteristics and simplifying computational demands in an interpretable way. Thus, \ac{sr} streamlines the estimation process, as well as provides interpretable, symbolic models suitable for real-time and resource-constrained applications inherent in the next-generation communication systems. We focus on \ac{aoa} estimation with a single antenna at the receiver; transmitter-side beamforming and \ac{ris} are part of the link. Moreover, the receiver provides only a scalar S-parameter per frequency. 
 
The contributions of this paper are as follows:
\begin{itemize}

  \item We design a two‐stage measurement that establishes an experimental foundation to support the proposed analytical framework. The first stage performs fine-grained sweeps in an anechoic chamber across azimuth angles and frequency bands to characterize the free-space beam pattern. The second stage measures a realistic indoor \ac{ris}-aided scenario using a passive \ac{ris} with fixed incidence geometry. This two-step setup validates the closed-form mapping between path loss and beam pattern under both idealized and practical conditions and provides a reproducible template for future mmWave and \ac{ris} studies.
  \item We propose \ac{sr}‐based \ac{ml} framework: SABER to extract interpretable, closed‐form models for both beam pattern and \ac{aoa} estimation from path loss coefficient. As shown in the results, we can achieve sub-0.5 degree accuracy in the first stage (anechoic chamber). Moreover, in the second stage (\ac{ris}-aided system), our methods can recover the fixed \ac{aoa} with near error-free performance. Furthermore, we show that by injecting just a small amount of prior structure, we obtain models that retain essentially the same level of accuracy but remain fully transparent and physically meaningful, whereas a purely unconstrained \ac{sr} fit may arrive at better performance, ($0.396^\circ$ vs $0.42^\circ$ in Stage~I) at the cost of non‐intuitive expressions and no guarantees on interpretability. 
  
  \item We benchmark SABER against the theoretical estimation lower bound: \ac{crlb}. The results demonstrate that our interpretable inversions are near the \ac{crlb}. For example, in the $50^\circ$ to $60^\circ$ region, SABER is on the order of $10^{-3}$ degrees above its \ac{crlb} in Stage~II, while Stage~I is essentially indistinguishable from its bound. Elsewhere across the field of view, in both stages, track the \ac{crlb} closely, with noticeable deviation only as the angle approaches $90$ degrees. 
    \item We adopt an antenna-design perspective to derive closed‐form expressions that map path loss coefficients to beam‐pattern characteristics, enabling accurate beam‐pattern approximation and further gain precise \ac{aoa} estimations without the need for extensive snapshot collection. By parameterizing the main‐lobe response as a $\cos^n(\theta)$ model, we show how measured path loss coefficients can be inverted to recover the incident signal direction with \ac{mae} of $0.42^\circ$, while the unconstrained method obtains accuracy of $0.396^\circ$.
  
\end{itemize}

\noindent The paper is organized as follows: Section II introduces the mathematical framework for path loss-based \ac{aoa} estimation. Section III details the proposed Symbolic Regression (SR)-based framework, SABER, and its workflow. Section IV describes the two experimental setups—a controlled free-space measurement and a \ac{ris}-aided indoor testbed—along with the data collection procedures , while Section V covers the model training parameters. Section VI presents the results, evaluating the SR-based methods and comparing their performance to the theoretical \ac{crlb}. Finally, Section VII concludes the paper and outlines future research directions.



\section{Analytical Model for Path Loss to Beam Pattern Mapping}
\label{sec:meth_math}

In this section, we describe two scenarios to demonstrate the method for calculating the total path loss coefficient during measurement in an indoor environment. We first consider a scenario in which the transmission takes place in a free space with far-field setup, and then a scenario in which a connection is established via \ac{ris} with no direct link between transmitter and receiver, and the total path loss is calculated as the sum of path loss between the transmitter and the receiver, the antenna gains/beamforming and the losses during the measurement. Consequently, the path loss coefficient can be defined as follows:

\begin{equation}
    PL_{\_,\mathrm{tot}}=PL_{\_}+\sum_{loss} L_{loss}-\sum_{gain}G_{gain},
    \label{eq:path loss_00}
\end{equation}

\noindent where $PL_{\_}$ denotes as the path loss between the transmitter and the receiver for the specific scenario, i.e., FS for free space and RIS for RIS-aided scenario. 

\subsection{Free space scenario}

\begin{figure}[t!]
    \centering
    \includegraphics[width=0.48\textwidth]{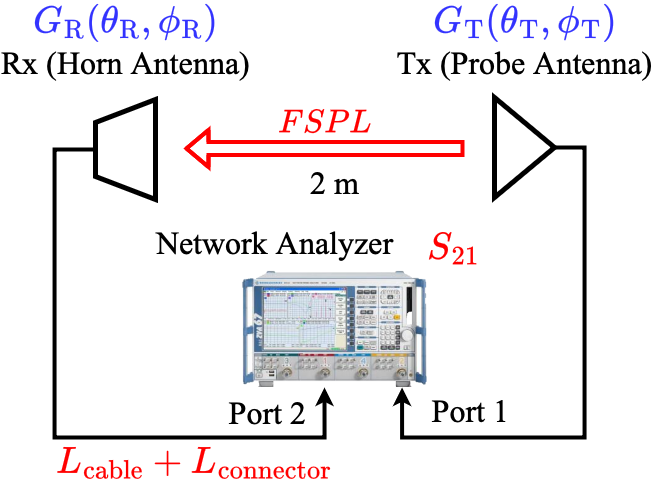}
    \caption{System model of free-space scenario.}
    \label{fig:systemmodel_chamber}
\end{figure}

In this scenario, the transmitter (Tx) and receiver (Rx) are in their far-field regions. Therefore, the path loss coefficient is modeled using the \ac{fspl} equation, denoted as $PL_{\mathrm{FS}}$. It is defined as:

\begin{equation}
\begin{aligned}
    PL_{\mathrm{FS}}~[\mathrm{dB}]=&20\log_{10}(R)+20\log_{10}(f)\\
    &+20\log_{10}(4\pi)-20\log_{10}c,
\end{aligned}
    \label{eq:FSPL_00}
\end{equation}

\noindent where $R$, $f$, and $c$ are the distance between transmitter and receiver, the carrier frequency of the signal, and the speed of light, respectively. As depicted in the Fig~\ref{fig:systemmodel_chamber}, during the measurement, the losses of the connector and the cable, denoted as $L_{\mathrm{connector}}$ and $L_{\mathrm{cable}}$, respectively, along with the gain of the transmitting and receiving antennas, $G_{\mathrm{T}}$ and $G_{\mathrm{R}}$, are taken into account. As a result, the total path loss coefficient can be expressed as follows:

\begin{equation}
\begin{aligned}
    PL_{\mathrm{FS,tot}}~[\mathrm{dB}] = &PL_{\mathrm{FS}}+L_{\mathrm{connector}}+L_{\mathrm{cable}}\\
    &-G_{\mathrm{T}}\left(\theta_{\mathrm{T}},\phi_{\mathrm{T}}\right)-G_{\mathrm{R}}\left(\theta_{\mathrm{R}},\phi_{\mathrm{R}}\right).
\end{aligned}
        \label{eq:fspl}
\end{equation}

\noindent Throughout the work, we assume the losses of connector and cable to be $1$~dB each. Moreover, the terms 
$G_{\mathrm{T}}\left(\theta_{\mathrm{T}},\phi_{\mathrm{T}}\right)$ and $G_{\mathrm{R}}\left(\theta_{\mathrm{R}},\phi_{\mathrm{R}}\right)$ are defined as the gains of the transmitting and receiving antennas, respectively, as functions of the azimuth ($\theta_{\_}$) and elevation ($\phi_{\_}$) angles in the spherical coordinates. 

\begin{figure}[b!]
    \centering
    \includegraphics[width=0.37\textwidth]{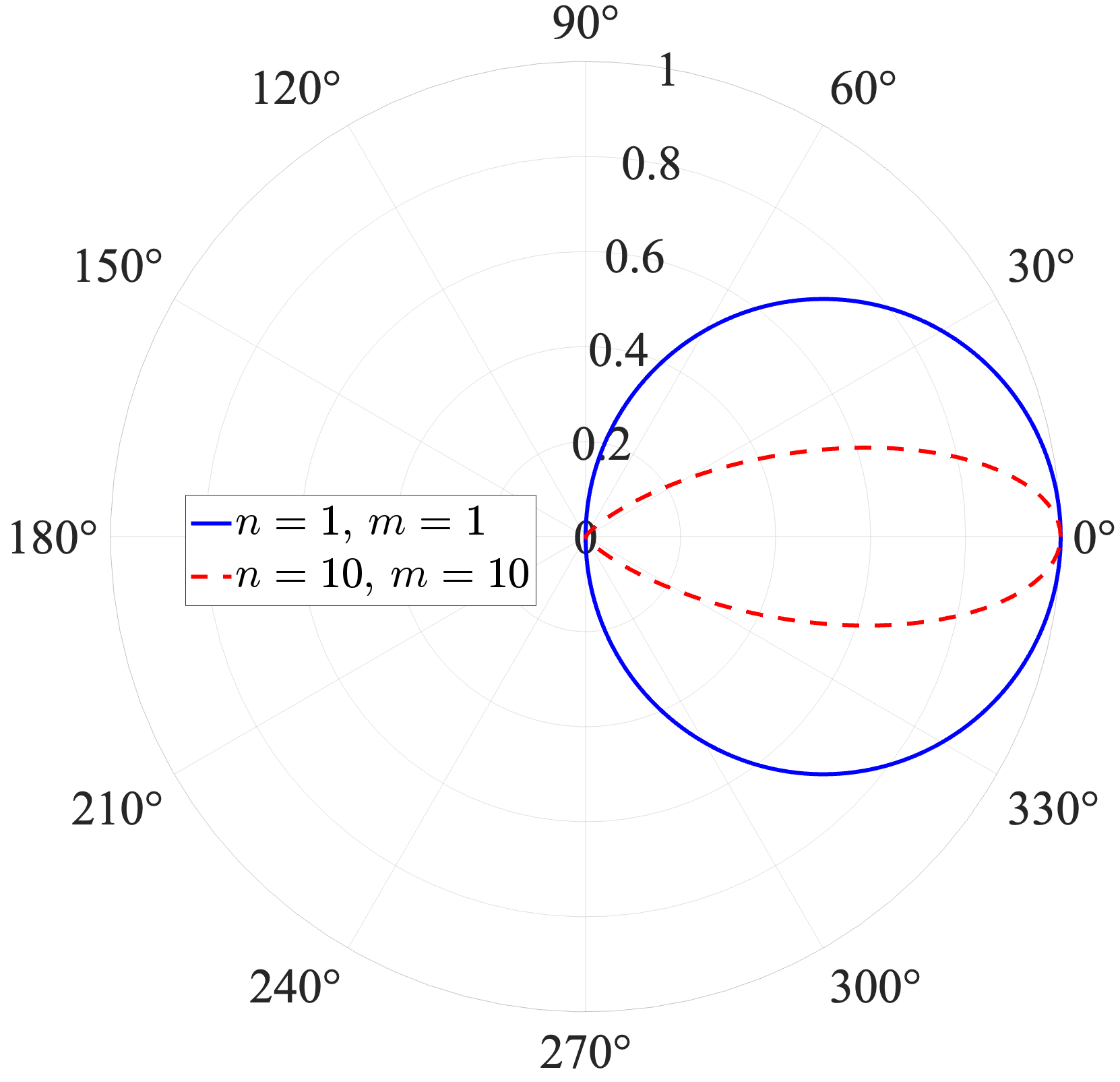}
    \caption{Beam pattern with different $n$ and $m$ in polar coordinates.}
    \label{fig:beampattern_nm}
\end{figure}

For simplicity, we assume that both the transmitting and receiving antennas exhibit the same directional characteristics. Specifically, the gain of each antenna is modeled as the product of its maximum gain, $G_{\max}$, and its normalized radiation pattern, $U(\theta, \phi)$. Thus, the antenna gain $G(\theta, \phi)$ is given by:

\begin{equation}
    G\left(\theta,\phi\right) = G_{\max} U\left(\theta, \phi\right).
    \label{eq:gain_def0}
\end{equation}

\noindent Moreover, the normalized radiation pattern $U(\theta, \phi)$ can be modeled as:

\begin{equation}
\begin{aligned}
    U\left(\theta, \phi\right) = \cos^{n}\left(\theta\right) &\cos^{m}\left(\phi\right),\\
    &\mathrm{where}~\ \theta \in [0, \frac{\pi}{2}], \ \phi \in [-\pi, \pi],
\end{aligned}
    \label{eq:beam_pattern}
\end{equation}

\noindent where $n$ and $m$ are the parameters determining the directivity of the antenna and the shape of the pattern~\cite{why_cosine}. In particular, as the values of $n$ and $m$ increase, the main lobe becomes narrower, resulting in a more focused and directional beam. This behavior is illustrated in Fig.~\ref{fig:beampattern_nm}, which shows the azimuth angle of the beam pattern for two sets of parameters. A small $n$ and $m$ produce a wider, less directional beam, whereas larger values have a narrower and higher directive beam. As a result, by substituting Eq.~\eqref{eq:fspl}, Eq.~\eqref{eq:gain_def0}, and Eq.~\eqref{eq:beam_pattern} in Eq.~\eqref{eq:path loss_00}, the total \ac{fspl}  can be rewritten as:

\begin{equation}
    \begin{aligned}
        PL_{\mathrm{FS,tot}}&~[\mathrm{dB}] = PL_{\mathrm{FS}}+L_{\mathrm{connector}}+L_{\mathrm{cable}} \\
        &\quad - G_{\mathrm{T},\max} - G_{\mathrm{R},\max} \\
        &\quad - 10\left(n_{\mathrm{T}}\log_{10}(\cos(\theta_{\mathrm{T}})) + m_{\mathrm{T}}\log_{10}\right.(\cos(\phi_{\mathrm{T}})) \\
        &\left.\quad + n_{\mathrm{R}}\log_{10}(\cos(\theta_{\mathrm{R}})) + m_{\mathrm{R}} \log_{10}(\cos(\phi_{\mathrm{R}}))\right).
    \end{aligned}
    \label{eq:path_loss_dB_theta_phi}
\end{equation}

\subsection{RIS-aided Systems}
\begin{figure}[t!]
    \centering
    \includegraphics[width=0.48\textwidth]{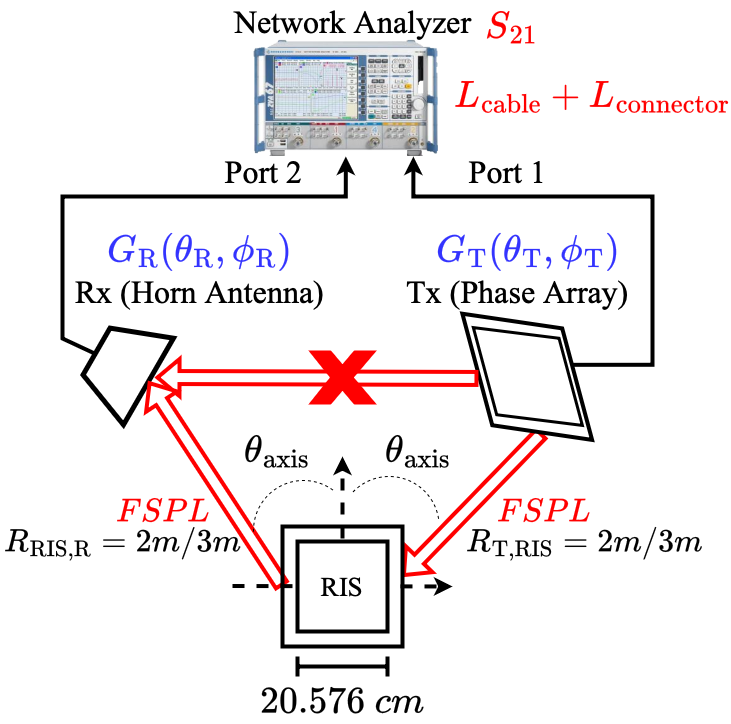}
    \caption{System model of RIS-aided system.}
    \label{fig:systemmodel_RIS}
\end{figure}

\begin{figure*}[t!]
        \centering
    \includegraphics[width=1\textwidth]{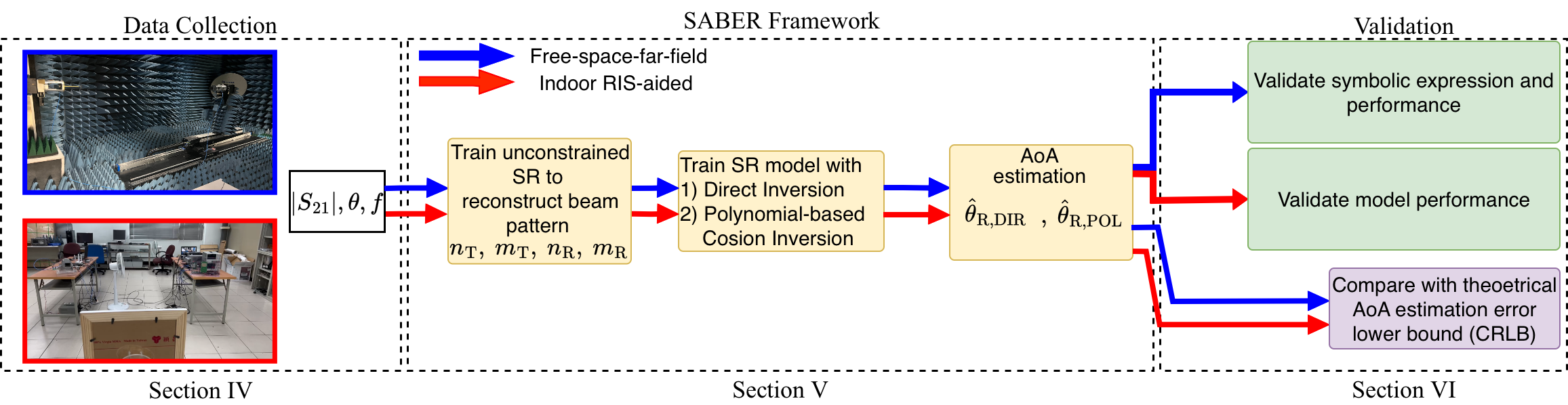}
    \caption{Flow chart of SABER: from the data collection to verification.}
    \label{fig:flowchart}
\end{figure*}
The system model for a \ac{ris}-aided system is shown in Fig.~\ref{fig:systemmodel_RIS}. According to~\cite{RIS_PL}, the path loss coefficient, $PL_{\mathrm{RIS}}$, for a RIS-aided system from transmitter to receiver via the intelligent surface in free space and within the far-field regime is given by:

\begin{equation}
\begin{aligned}
     PL_{\mathrm{RIS}}=\frac{G_{\mathrm{T}}G_{\mathrm{R}}}{\left(4\pi\right)^2}&\left(\frac{ab}{R_{\mathrm{T,RIS}}R_{\mathrm{RIS,R}}}\right)^{2}\varepsilon^{2}_{\mathrm{ap}}\cos^{2}\left(\theta_{\mathrm{
    axis}}\right)\\ 
    &\quad \quad \quad \quad,~ 0\leq \theta_{\mathrm{axis}}\leq\frac{\pi}{2},
\end{aligned}
   \label{eq:PL_RIS}
\end{equation}

\noindent where $a$, $b$, $R_{\mathrm{T,RIS}}$ and $R_{\mathrm{RIS,R}}$ are the length, the width of the \ac{ris}, the distances from the transmitter to the \ac{ris} and the \ac{ris} to the receiver, respectively. In addition, $\varepsilon_{\mathrm{ap}}$ is the aperture efficiency of the \ac{ris} and $\theta_{\mathrm{axis}}$ is the angle between the incident wave and the optical axis of the RIS. This implies that the expression for the path loss from the source to the \ac{ris} is maximized when the transmitter is on the optical axis, i.e., when $\theta_{\mathrm{axis}}=0$. Similar to the free-space scenario, we also consider the losses of the connector ($L_{\mathrm{connector}}$), the transmission line ($L_{\mathrm{cable}}$), and the power divider ($L_{\mathrm{PD}}$). Particularly, the power divider has 4 layers, each layer introduces a $7.5$~dB signal loss. In addition to the gain of the transmitter and receiver, there is a beamforming gain, $G_{\mathrm{BF}}$, presents due to the nature of the phase array. Therefore, by substituting Eq.~\eqref{eq:gain_def0} into Eq.~\eqref{eq:PL_RIS}, we can then rewrite Eq.~\eqref{eq:path loss_00} in decibels as follows:

\begin{equation}
     \begin{aligned}
        PL_{\mathrm{RIS,tot}}&~[\mathrm{dB}] =10\left(\log_{10}G_{\mathrm{T}}(\theta_{\mathrm{T}},\phi_{\mathrm{T}})+\log_{10}G_{\mathrm{R}}(\theta_{\mathrm{R}},\phi_{\mathrm{R}})\right)\\
        &+20\left(\log_{10}a +\log_{10}b-\log_{10}R_{\mathrm{T,RIS}}\right.\\
        &-\log_{10}R_{\mathrm{RIS,R}}+\log_{10}\varepsilon_{\mathrm{ap}}+\left.\log_{10}\cos\theta_{\mathrm{axis}}\right)\\
        &+L_{\mathrm{PD}}+L_{\mathrm{connector}}+L_{\mathrm{cable}}-G_{\mathrm{BF}}.
    \end{aligned}
    \label{eq:PL_RIS_db}
\end{equation}

Building upon the analytical path loss models established in this section, we now introduce the proposed SR-based framework designed to empirically learn these relationships from measurement data.

\section{Proposed constrained SR-based ML Framework: SABER}
\label{sec:SRframework}
We adopt \ac{sr} to learn closed-form, interpretable relations between the measured path loss coefficient, beam pattern, and the \ac{aoa}. \ac{sr} has several major families. Evolutionary \ac{sr} via \ac{gp} refines expression trees through mutation, crossover, and selection; it remains a dominant approach because it can discover compact, human-readable formulas, though it can be compute-intensive in very large operator spaces~\cite{anaqreh2025automatedmodelingmethodpathloss}. \ac{dsr} uses an \ac{rnn} policy trained with \ac{rl}~\cite{hayes2025deepsymbolicoptimizationreinforcement} (e.g., \ac{rspg}, \ac{pqt}) to generate expressions rewarded by fit, typically with normalized error–based rewards and search constraints to keep formulas meaningful. To overcome the downsides of pure \ac{gp} or pure \ac{dsr}, hybrid, neural-guided \ac{gp} \ac{sr} is proposed with neural proposals and priority queues to accelerate exploration~\cite{mundhenk2021symbolicregressionneuralguidedgenetic}. Other lines include grammar/latene-variabl \ac{sr} (e.g., GrammarVAE)~\cite{kusner2017grammarvariationalautoencoder} and physics-inspired heuristics (AI Feynman)~\cite{udrescu2020aifeynmanphysicsinspiredmethod}, which help with syntax or exploit symmetries but may struggle to recover exact formulas or do not perform a direct symbolic search. 

Given our dataset size and the need for closed-form, physics-consistent inversions, we adopt \ac{gp}-based \ac{sr} for SABER. \ac{gp} supports (i) parsimony via complexity penalties and Pareto selection, (ii) light domain priors that bias the search toward physically valid forms, and (iii) constant-time inference once the expression is found, which are the properties well aligned with timely \ac{aoa} estimation. Moreover, Fig.~\ref{fig:flowchart} outlines the workflow and paper organization: we first describe how we conduct data collection in Section~\ref{sec:meth_measure} (as shown in the first block of Fig.~\ref{fig:flowchart}), followed by the proposed SABER framework in Section~\ref{sec:modeltrainingandval} (highlighted in yellow), including training the \ac{sr}-based method to reconstruct the beam patterns and estimate the \ac{aoa}. Next, we validate the learned symbolic expressions and their predictive performances (marked in green). Finally, we benchmark against the \ac{crlb} to showcase how close they are to the theoretical lower bound on estimation error (marked in purple).   



\section{Data collection for model training}
\label{sec:meth_measure}

In this section, we describe the measurement schemes used to train the proposed \ac{sr}-based \ac{ml} models, corresponding to the blue-highlighted block in Fig.~\ref{fig:flowchart}. We collect measurement data from two real-world scenario, and utilize the collected data to train \ac{sr}-based \ac{ml} models and further validate their performance. The experimental validation directly follows the two‐stage scenario introduced in Section II: Stage I, depicted in Fig.~\ref{fig:systemmodel_chamber}, corresponds to free‐space beam‐pattern inversion in an anechoic chamber, we aim to find out if the \ac{sr} models can obtain closed-form expression and evaluate their performance, while Stage~II (shown in Fig.~\ref{fig:systemmodel_RIS}) applies the same \ac{sr}–based methods to a RIS‐aided indoor scenario, and see if the \ac{aoa} can be recovered.

\begin{figure}[t!]
    \centering
    \includegraphics[width=0.48\textwidth]{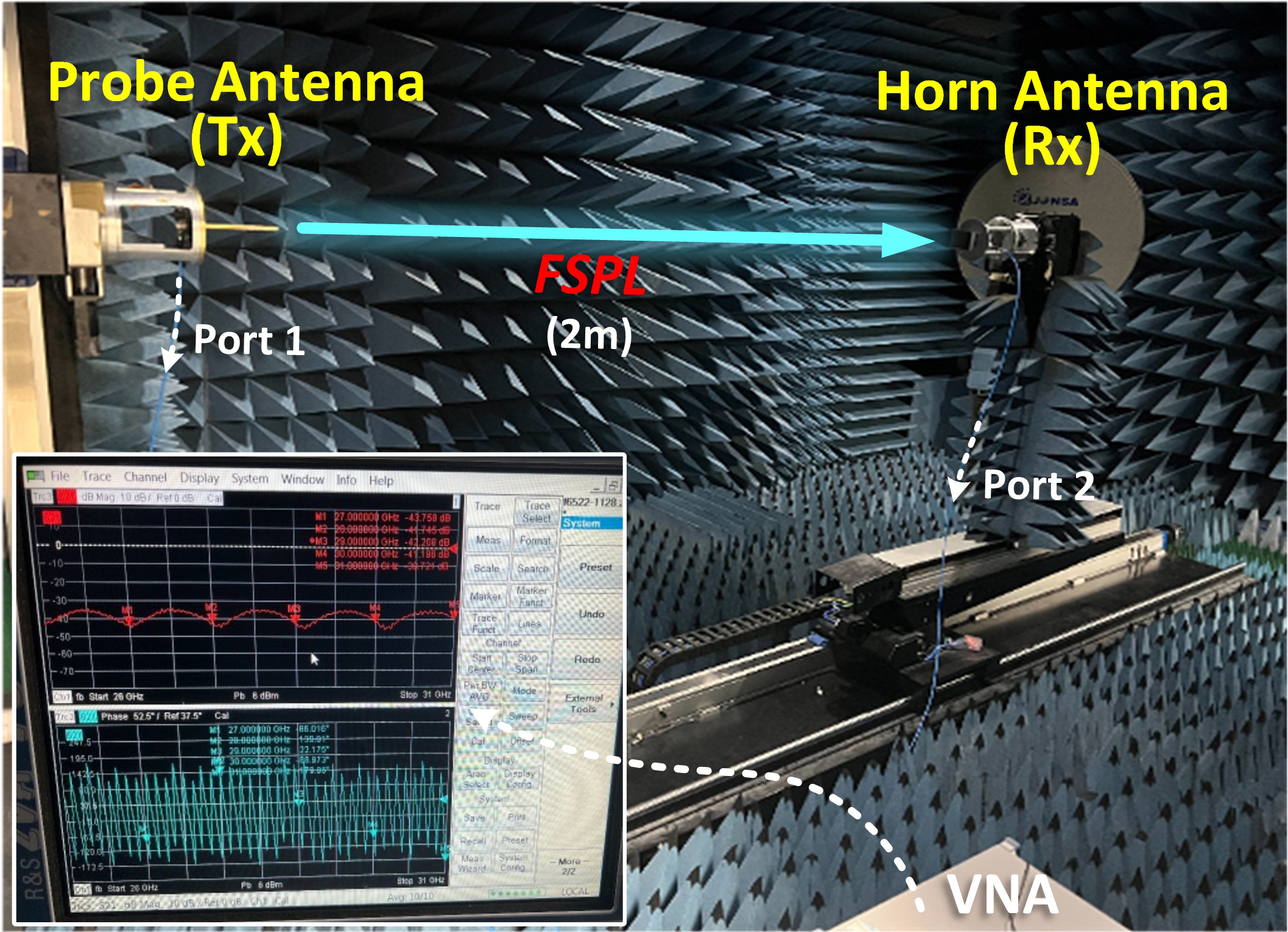}
    \caption{Experiment setup for the stage I (anechoic chamber verification). The probe antenna (Tx) is mounted on a motorized rotator at the left and the horn antenna (Rx) is fixed at the right, forming a 2 m free-space path (FSPL). Ports 1 and 2 of a Rohde \& Schwarz ZVA40 VNA connect to the Tx and Rx, respectively, to measure the forward transmission coefficient $S_{21,\mathrm{FS}}$ across 26 to 31 GHz. A representative sweep is shown in the lower-left box. }
    \label{fig:measurementsetup}
\end{figure}


\begin{figure}[t!]
    \centering
    \includegraphics[width=0.48\textwidth]{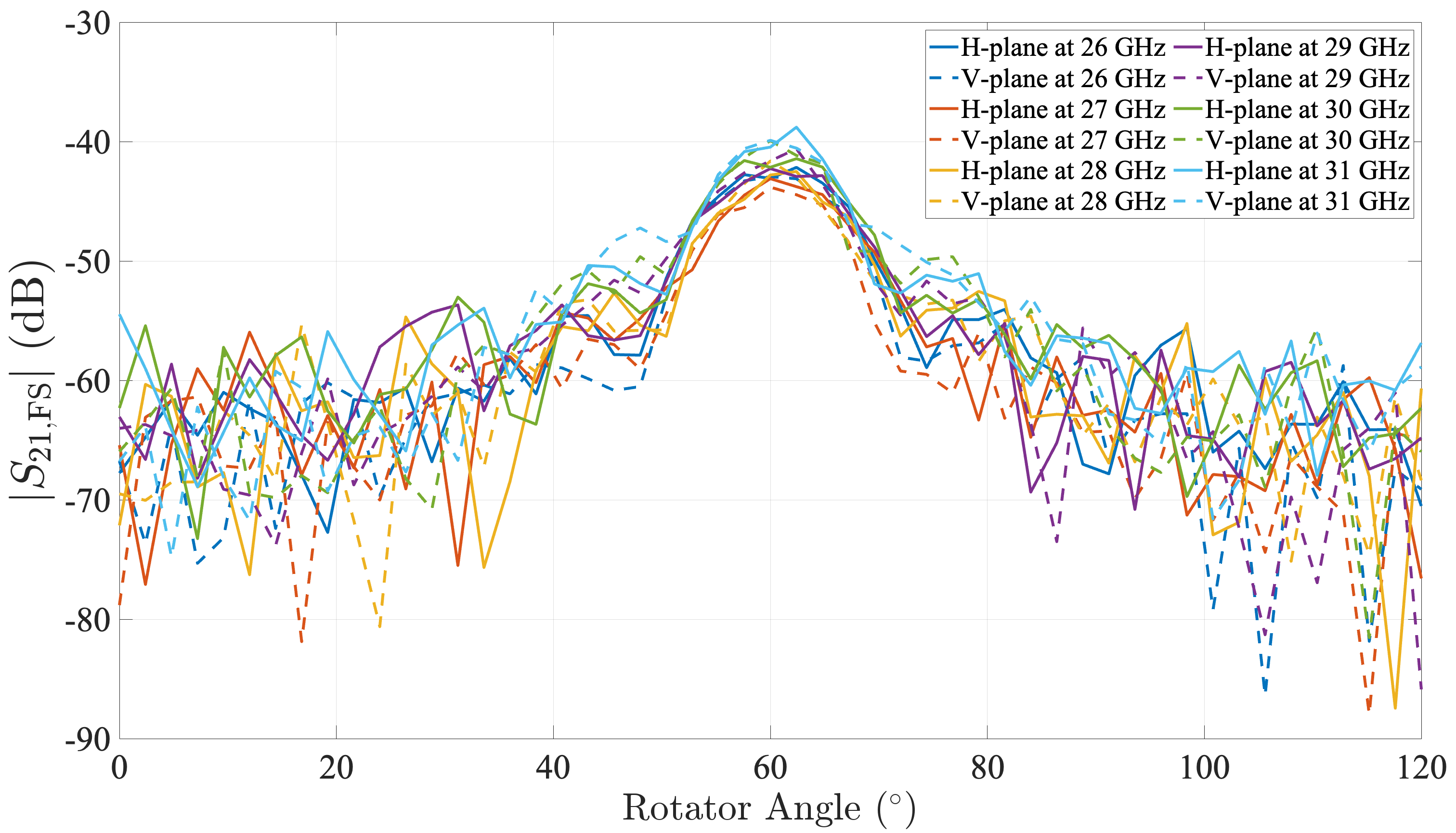}
    \caption{Measurement with Rohde \& Schwarz ZVA40 VNA for the forward transmission coefficient $S_{21,\mathrm{FS}}$ across 26 to 31 GHz.}
    \label{fig:thereal_measurementsetup_vna}
\end{figure}

\subsection{Stage I: anechoic chamber verification}

\label{sec:meth_measure_FSPL}
We first quantify system behavior in a controlled environment to extract beam-pattern response. To accurately quantify the \ac{fspl} between two antennas, a Rohde \& Schwarz ZVA40 Vector Network Analyzer (VNA)~\cite{RohdeSchwarz2020} is utilized. The experiments are carried out in a $7 \times 4 \times 3$~$\mathrm{m}^{3}$ anechoic chamber under far-field conditions. The antennas are arranged at a fixed distance of $2$~m, whereby the measurements are conducted across the frequency band from $26$ to $31$~GHz. In addition, the experimental configuration is depicted in Fig.~\ref{fig:measurementsetup}, illustrating the spatial arrangement of the antennas within the anechoic chamber. The transmitting antenna is a probe antenna placed on the left side of the figure, while a rotatable platform is placed on the right side of the figure, and the receiving antenna is a horn antenna. This platform allows angular rotation from \ang{0} to \ang{120}. In particular, at a rotation angle of \ang{60}, the antennas are directly facing each other, which optimizes signal reception and ensures maximum coupling efficiency. The probe (Tx) and horn antennas (Rx) have gains of $4.5$~dBi and $23.5$~dBi, respectively. 

Fig.~\ref{fig:thereal_measurementsetup_vna} shows the measured S-parameter denoted as $S_{21,\mathrm{FS}}$, expressed in~dB, which characterizes the forward transmission coefficient of the system under test. Both horizontal (H-plane) and vertical (V-plane) polarization are measured. This parameter, critical for assessing signal attenuation, is recorded over a frequency range of $26$~GHz to $31$~GHz in $1$~GHz intervals, with measurements taken at incremental angular positions from \ang{0} to \ang{120} in \ang{2.4} steps, in total of $51$ scanning points per frequency. The corresponding system model, illustrating the geometry and key parameters, is shown in Fig.~\ref{fig:systemmodel_chamber}. The data reveals that the maximum transmission, indicated by the peak $S_{21,\mathrm{FS}}$ values, occurs around \ang{60} to \ang{62.4} for both H- and V-planes, corresponding to the direct alignment of the antennas. The $S_{21,\mathrm{FS}}$ values across the dataset range from approximately $-88$ ~dB to $-39$ ~dB, illustrating the variation in signal strength with respect to frequency and angular position. 

\begin{figure}
\centering
\includegraphics[width=0.48\textwidth]{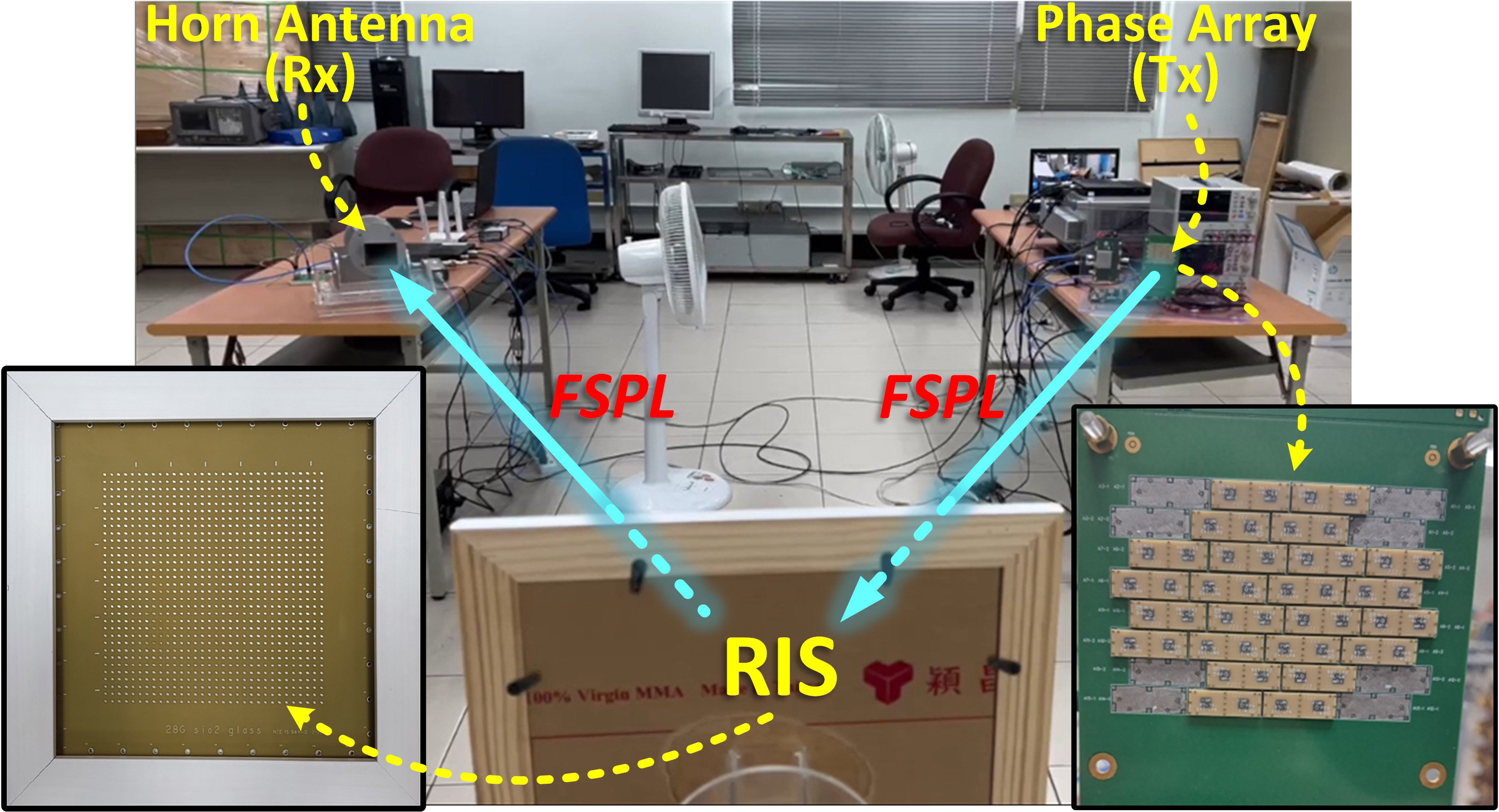}
\caption{Experimental setup for Stage II (RIS-aided scenario). View from the RIS side: a 48-element phase array transmitter (right inset, phase array~\cite{RIS_phasearray}) illuminates a passive 1024-element RIS (center)~\cite{RIS_passive_ref}, which reflects toward a horn-antenna receiver located at the left side. The Tx–RIS and RIS–Rx links are free-space line-of-sight (FSPL), as indicated by the arrows.}
\label{fig:systemmodel_RIS_aided}
\end{figure}

\subsection{Stage II: Real-world RIS-Aided indoor system verification}
\label{sec:meth_measure_RIS}



Building on the previous stage, the second stage is set in a realistic \ac{ris}-aided indoor testbed to verify that the measured S-parameter, $S_{21,\mathrm{RIS}}$, can be used to recover the \ac{aoa}. The experiment setup is  shown in Fig.~\ref{fig:systemmodel_RIS_aided}. Specifically, the transmitter is equipped with a 48-element phase array~\cite{RIS_phasearray}, shown in the the right inset, controlled by a beamformer (Renasas F6522~\cite{renesas_f6522}), while the receiver is equipped with a horn antenna. In addition, a \ac{ris} with passive elements is set up to relay the signal from the transmitter to the receiver. The front side of the \ac{ris} is shown in the left inset of Fig.~\ref{fig:systemmodel_RIS_aided}, the deployed \ac{ris} derived from the design in~\cite{RIS_passive_ref}, which has a diameter of $20.576\times 20.576$~cm$^{2}$ and is implemented using truncated patch antennas printed on a $1$ mm thick FR-4 substrate; the total number of patch antennas is $1024$. To achieve beam steering, we employ patch antennas with a shorted or opened load to simulate the ON/OFF status of a PIN diode, which is sequentially controlled column by column in the array. This configuration enables a progressive phase delay of $180^{\circ}$ for beam steering in the azimuthal direction, denoted by $\theta_{\mathrm{axis}}$. In addition, the steering angle is determined by the element spacing $d$ and the wavelength of the incident wave $\lambda$, which is given by:

\begin{equation}
\theta_{\mathrm{axis}}=\sin^{-1}\Bigl(\frac{\lambda}{2d}\Bigr).
  \label{eq:theta_s}
\end{equation}

\begin{figure}[t!]
    \centering
    \includegraphics[width=0.48\textwidth]{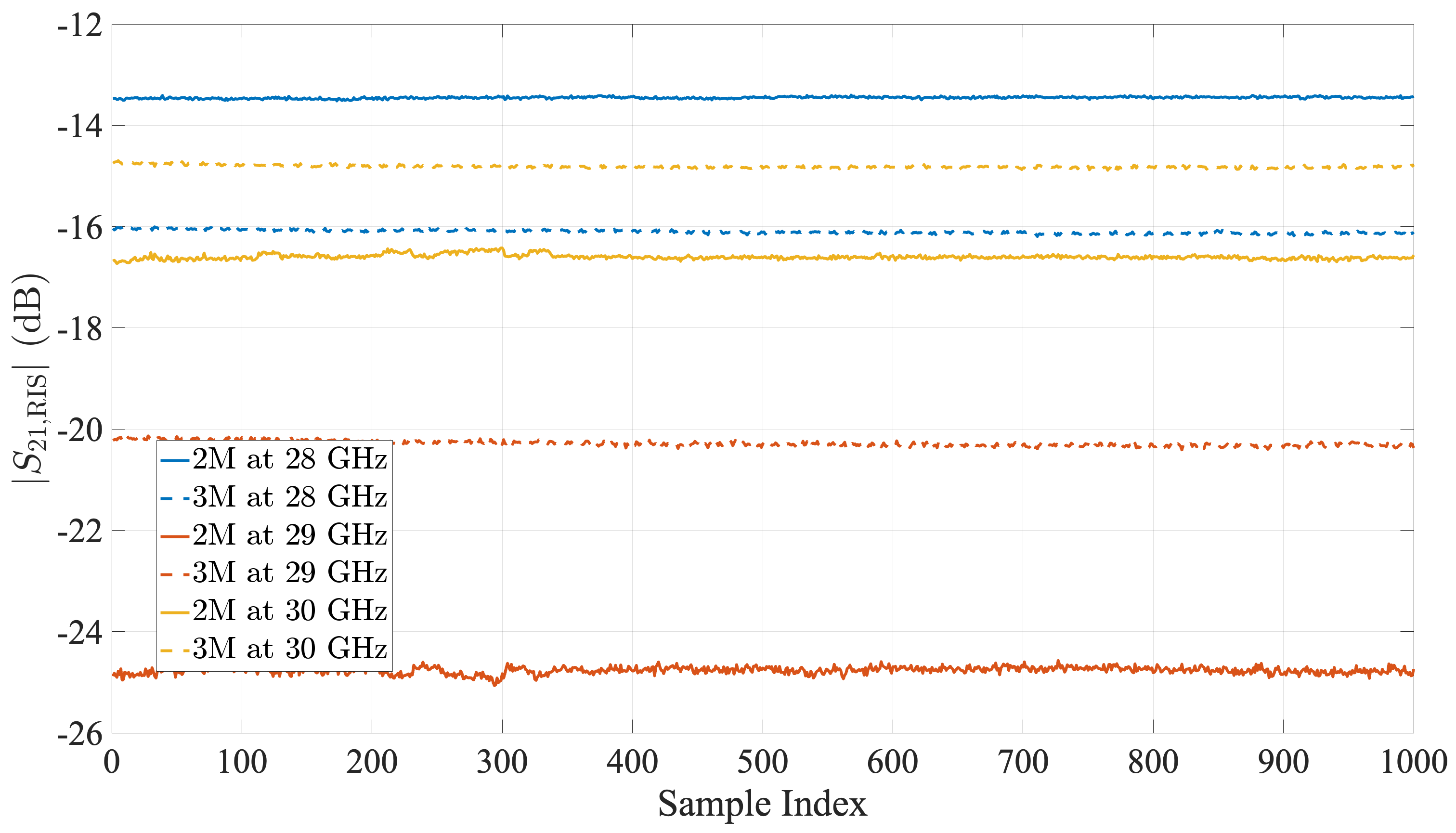}
    \caption{Measurement for the forward transmission coefficient, $S_{21,\mathrm{RIS}}$ in Stage II: RIS-aided indoor system.}
    \label{fig:measurement_RIS}
\end{figure}

\noindent In addition, the experiments are conducted in two different distances, the detailed parameters are also shown in Fig~\ref{fig:systemmodel_RIS}. First, we set the distances between \ac{ris} to both of the transmitter and the receiver as $2$~m. Then the distances are increased to $3$~m, we denote these two measurements as $S_{21,\mathrm{RIS,2m}}$ and $S_{21,\mathrm{RIS,3m}}$, respectively. The measurement can be seen in the Fig~\ref{fig:measurement_RIS}. For the sake of simplicity, we set the angle between both transmitter and receiver to $70^{\circ}$, i.e., $\theta_{\mathrm{axis}}=35^{\circ}$. Since the distance between the transmitter and the RIS as well as between the RIS and the receiver is known, we can obtain the \ac{aoa} is $55^\circ$ with simple geometric manipulations. Furthermore, due to the fact that the transmitter is equipped with a beamformer, there is \ac{nlos} connection between the transmitter and receiver, resulting in negligible signal strength. The S-parameters are measured in the 28, 29, and 30 GHz frequency bands. The measurement results reveal that $S_{21,\mathrm{RIS,3m}}$ at 28 GHz has a mean attenuation of -16.08 dB, which is approximately 2.6 dB higher than the mean value for $S_{21,\mathrm{RIS,2m}}$. However, this trend is not persist across the other frequencies, which emphasises the dominant, frequency-selective role of the \ac{ris}. Notably, a significant signal degradation is observed at 29 GHz for the $2$~m configuration, where the mean value of $S_{21,\mathrm{RIS,2m}}$ drops to $-24.76$ dB, far weaker than the $-20.26$ dB of $S_{21,\mathrm{RIS,3m}}$, indicating the formation of a deep, destructive interference null for this geometry. Conversely, at $30$ GHz,the $3$~m link is superior, with its mean $S_{21,\mathrm{RIS,3m}}$ of $-14.81$ dB outperforming the $-16.61$ dB of $S_{21,\mathrm{RIS,2m}}$. To summarize, the strong dependence of the measured S-parameters on both frequency and geometry serves as compelling experimental validation, confirming the fundamental ability of \ac{ris} to shape the electromagnetic field by strategically creating complex patterns of constructive and destructive interference~\cite{RIS_measure}.

\section{Development of SABER}
\label{sec:modeltrainingandval}
As shown in Fig.~\ref{fig:flowchart}, data collection is followed by model training and validation, which are respectively highlighted in red and green blocks. We train \ac{sr} models with open-source tool: PySR~\cite{cranmer2023interpretablemachinelearningscience}. In addition, the unconstrained \ac{sr} is considered as the baseline estimator, which learns a free-form mapping $g_{\mathrm{SR}}(\cdot)$. It can be expressed as follows:

\begin{equation}
    \hat{\theta}_{\mathrm{R,SR}}=g_{\mathrm{SR}}(|S_{21,\_}|_{\mathrm{lin}}). 
    \label{eq:theta_SR}
\end{equation}

\noindent where $|S_{21}|_{\mathrm{lin}}$ denotes the measured S-parameter in linear scale. Moreover, expression trees are built using unary functions includes $\left\{\cos(\cdot), \sin(\cdot), \log_{10},  |\cdot|\right\}$, while the mathematical operations are limited to addition, subtraction, multiplication and division. In addition, we set the number of evolutionary iterations to $5000$ and the maximum depth to $10$, and Pareto selection on error–complexity. In particular, we define the differential path loss, $\Delta PL$, which is calculated as:

\begin{equation}
    \Delta PL = PL_{\mathrm{\_,tot}}-PL_{\mathrm{\_}}.
    \label{eq:deltaPL}
\end{equation}

The proposed framework first applies unconstrained \ac{sr} to extract beam-pattern parameters $n_{\mathrm{T}}$, $m_{\mathrm{T}}$, $n_{\mathrm{R}}$, and $m_{\mathrm{R}}$. Subsequently, two physics-guided estimation methods: direct inversion and polynomial-based Cosine inversion are proposed for estimating \ac{aoa}:

\subsection{SABER: Direct Inversion}

In this approach, we leverage the analytically known cosine-shaped antenna beam pattern to reverse the mapping from normalized path loss to \ac{aoa}. Specifically, we fix the receive directivity ($n_{\mathrm{R}}$) and fit only a single additive offset is injected into the model to obtain the angle, denoted as $\hat\theta_{\mathrm{R,DIR}}$, and it is expressed as:

\begin{equation}
    \hat{\theta}_{\mathrm{R,DIR}}= \mathrm{arccos}\left(10^{|S_{21,\_}|}\right)+\mathrm{offset}.
    \label{eq:theta_dir}
\end{equation}

\subsection{SABER: Polynomial-based Cosine Inversion}

Inspired by the fact that cosine function can be well approximated by low-order polynomials, in this approach, we direct fit a quadratic surrogate for $\cos{(\hat{\theta}_{\mathrm{R}}})$, i.e., $\cos\theta \;\approx\; a\,(\Delta PL)^2 + b\,(\Delta PL) + c$, and then we can obtain the angle with $\arccos{(\cdot)}$. As a result, the estimated angle can then be expressed as: 

\begin{equation}
    \hat{\theta}_{\mathrm{R,POL}}= \arccos \left(10^{\frac{a\,(\Delta PL)^2 + b\,(\Delta PL) + c}{10n_{\mathrm{R}}}}\right).
    \label{eq:theta_poly}
\end{equation}
To consolidate the methodologies described in this section, we present our complete framework in Algorithm~\ref{alg:saber}. The algorithm formalizes the key procedures for model training (Fit) and angle estimation (Predict). 

First, it begins by taking the measured path loss coefficient as input $PL_{\mathrm{meas}}$ (or equivalently, the magnitude of the scattering parameter $|S_{21}|$) together with the ground-truth \ac{aoa} labels $\theta_{\mathrm{R}}$. The algorithm can operate in one of two modes: Direct Inversion or Polynomial-based Cosine Inversion. A set of hyperparameters for the \ac{sr} engine (PySR) is also defined, including the number of iterations, maximum expression size, and the set of mathematical operators allowed during the symbolic search. The goal of the algorithm is to produce a closed-form estimator $g_{\mathrm{SR}}$ that maps path-loss $PL$ to the predicted \ac{aoa} ($\hat{\theta}_{\mathrm{R}}$). This is described in the required and ensure part of the algorithm and followed by the initialization part of the algorithm (labeled as line 1 to 4). SABER first computes the path loss difference $\Delta PL = PL_{\mathrm{meas}} - PL_{\mathrm{model}}$, which represents the deviation of the measured signal from the ideal theoretical model. It then initializes the radiation pattern of the antenna ($U(\theta, \phi)$) and fits the parameters that define its shape, namely $G_{\max}$, $\theta_{\mathrm{R}}$, and $\theta_{\mathrm{T}}$. From this, it extracts  $n/m_{\mathrm{R}}$ and $n/m_{\mathrm{T}}$ that characterize the beam directivity for both the receiver and transmitter. The antenna gain is then converted into decibel scale as $G_{\mathrm{dB}} = 10\log_{10} G(\theta, \phi)$, preparing the data for SABER. Conceptually, this step corresponds to the SABER framework in the Fig.~\ref{fig:flowchart}: first yellow block, where the reconstruction of the beam pattern parameters from measurement data happens. 

Next, the algorithm proceeds according to the selected mode. If the Direct Inversion mode is chosen, SABER defines an estimator based on the analytical inversion of the cosine relationship between path loss and incident angle. Specifically, it uses $g_{\mathrm{SR}}(x) = \arccos(10^{x/(10n_R)}) + \text{offset}$. PySR is then used to fit this expression and refine the result through symbolic regression. If the Polynomial-based Cosine Inversion mode is chosen instead, the algorithm defines a polynomial surrogate $q(x)$ to approximate the cosine term within its valid range of $[-1,1]$. The estimator is then expressed as $g_{\mathrm{SR}}(x) = \arccos(\min\{1,\max\{-1,q(x)\}\})$, ensuring that physical constraints are respected. PySR fits this surrogate polynomial, balancing the trade-off between model accuracy and expression complexity through Pareto optimization. This corresponds to the second yellow block in SABER framework in Fig.~\ref{fig:flowchart} and line 5 to 13 in Algorithm~\ref{alg:saber}. 

After training in either mode, the algorithm returns the learned symbolic model $g_{\mathrm{SR}}$, which serves as the closed-form mapping between path-loss measurements and \ac{aoa} (line 14 to 17 in Algorithm~\ref{alg:saber}). In practice, this model is used for inference on unseen data, which the algorithm defines in the predict function. During prediction, a new path loss coefficient  $PL_{\mathrm{test}}$ is passed to the trained estimator $g_{\mathrm{SR}}$, which outputs the predicted AoA $\hat{\theta}_{\mathrm{R}}$. This corresponds to the last yellow block “AoA Estimation” of SABER framework in Fig.~\ref{fig:flowchart}, where the outputs $\hat{\theta}_{\mathrm{R},\text{DIR}}$ and $\hat{\theta}_{\mathrm{R},\text{POL}}$ are produced.

Finally, although not part of the algorithm itself, these estimated angles are validated experimentally. The symbolic expressions are first analyzed for interpretability and physical consistency, and then their estimation performance is compared against measured ground truth as well as the theoretical lower bound given by the \ac{crlb}. This matches the validation section of Fig.~\ref{fig:flowchart}, where both the symbolic regression results and their numerical accuracy are evaluated.

\begin{algorithm}[t]
\caption{proposed SABER framework}
\label{alg:saber}
\footnotesize
\begin{algorithmic}[1]
\Require measured path loss $PL_{\mathrm{meas}}$ (or $|S_{21}|$), training labels $\theta_{\mathrm{R}}$;
        mode $m\in\{\textsc{DirectInv},\textsc{PolyCosInv}\}$;
        PySR hyperparameters: $n_{\mathrm{iter}}{=}5000$, $\mathrm{maxsize}{=}10$;
        operator sets: $\{\cos,\sin,\exp,\log_{10},|\cdot|,+,-,\times,\div\}$ 
\Ensure closed-form estimator $g_{\mathrm{SR}}: PL\mapsto \hat{\theta}_{\mathrm{R}}$
\Statex \textbf{Initialization:}
\State compute $\Delta PL \leftarrow PL_{\mathrm{meas}} - PL_{\mathrm{model}}$ \Comment{Eq.~\ref{eq:theta_dir}}
\State initialize radiation pattern $U(\theta, \phi)$
\State fit $(G_{\max},\theta_{\mathrm{R}},\theta_{\mathrm{T}})$ in $G(\theta,\phi)=G_{\max}\,U(\theta, \phi)$, find $n/m_{\mathrm{R}}$ and $n/m_{\mathrm{T}}$
\State set $G_{\mathrm{dB}} \leftarrow 10\log_{10} G(\theta,\phi)$
\If{$m=\textsc{DirectInv}$} \Comment{SABER: Direct Inversion}
  \State define the estimator for $x$:
        \[
          g_{\mathrm{SR}}(x)=\arccos\!\Big(10^{\frac{x}{10n_{\mathrm{R}}}}\Big)+\mathrm{offset}
        \]
  \State fit with PySR
\ElsIf{$m=\textsc{PolyCosInv}$} \Comment{SABER: Polynomial-based Cosine Inversion}
  \State define polynomial surrogate of cosine function $q(x)$
  \State define the estimator for $q(x)$:
        \[
          g_{\mathrm{SR}}(x)=\arccos\!\big(\min\{1,\max\{-1,\,q(x)\}\}\big)
        \]
  \State fit with PySR (Pareto selection: validation error vs.\ expression size)
\EndIf
\State \Return $g_{\mathrm{SR}}$
\Statex
\Function{Predict}{$PL_{\mathrm{test}},\,g_{\mathrm{SR}}$}
  \State $x \leftarrow PL_{\mathrm{test}}$
  \State \Return $g_{\mathrm{SR}}(x)$ 
\EndFunction
\end{algorithmic}
\end{algorithm}

The performance metrics in this work are \ac{mae} and \ac{rmse}, which quantify, respectively, the average absolute error and the root-mean-square error. For \(N\) samples with ground-truth angles \(\{\theta_i\}_{i=1}^N\) and estimates \(\{\hat{\theta}_i\}_{i=1}^N\), these two performance metrics are defined as:

\begin{equation}
\mathrm{MAE}=\frac{1}{N}\sum_{i=1}^{N}\left|\hat{\theta}_{i}-\theta_{i}\right|,
\label{eq:mae}
\end{equation}
\noindent and 
\begin{equation}
\mathrm{RMSE}=\sqrt{\frac{1}{N}\sum_{i=1}^{N}\left(\hat{\theta}_{i}-\theta_{i}\right)^{2}}.
\label{eq:rmse}
\end{equation}

\begin{table*}[htbp]
\centering
\caption{Summary of SR-based approaches, including the final interpretable expressions and the performance in MAE.}
\label{tab:summary_FS}
\begin{tabular}{@{}l S[table-format=2.2] l@{}}
\toprule
\textbf{Approach} & \textbf{MAE (\si{\degree})} & \textbf{Final interpretable expression} \\
\midrule
Unconstrained SR
  & 0.396
  & $\displaystyle
     \hat\theta_{\mathrm{R,SR}}
     = 0.53076\sin\!\bigl(\sin(0.12336\,PL_{\mathrm{FS,tot}} + \theta_{\mathrm{T}})\bigr)$ \\
SABER: Direct Inversion
  & 0.42
  & $\displaystyle \hat\theta_{\mathrm{R,DIR}}
     = \arccos \!\bigl(10^{\Delta PL/(10\,n_{\mathrm{R}})}\bigr)
     +0.05813$ \\
SABER: Polynomial-based Cosine Inversion
  & 5.96
  & $\displaystyle
     \begin{aligned}
       \hat\theta_{\mathrm{R,POL}} & \approx \arccos(0.03879\,\Delta PL^2
                  + 0.1165\,\Delta PL
                  + 0.8303)
     \end{aligned}$ \\
\bottomrule
\end{tabular}
\end{table*}

\section{Results and Discussions}
\label{sec:results}
In this section, we demonstrate the results of the unconstrained \ac{sr}-based and solutions and SABER for both beam pattern and \ac{aoa} estimations in the aforementioned scenarios and measurements. Our focus is on the azimuth angle, $\theta$, which is directly related to the values of $n_{\mathrm{T}}$ and $n_{\mathrm{R}}$.


\subsection{Stage I (anechoic chamber) results}
\label{subsec:bp}

\textit{Automatically learning to estimate the directivity parameters}: By applying the \ac{sr}-based method, we can determine all directivity-related parameters, which include $n_{\mathrm{T}}$, $n_{\mathrm{R}}$, $m_{\mathrm{T}}$, and $m_{\mathrm{R}}$. Our findings from this method indicate that $n_{\mathrm{T}}$ and $m_{\mathrm{T}}$ are $1$, while $n_{\mathrm{R}}$ and $m_{\mathrm{R}}$ are $28$. As a result, the expression in Eq.~\eqref{eq:path_loss_dB_theta_phi}, can be rewritten as:

\begin{equation}
    \begin{aligned}
        PL_{\mathrm{FS,SR}}~[\mathrm{dB}] &= FSPL  - G_{\mathrm{T},\max} - G_{\mathrm{R},\max} \\
        &\quad - 10\left(\log_{10}(\cos(\theta_{\mathrm{T}})) + \log_{10}\right.(\cos(\phi_{\mathrm{T}})) \\
        &\left.\quad + 280\cdot (\log_{10}(\cos(\theta_{\mathrm{R}})) + \log_{10}(\cos(\phi_{\mathrm{R}})))\right).
    \end{aligned}
    \label{eq:path_loss_dB_theta_phi_SR}
\end{equation}

 \begin{figure}[t]
    \centering
        \includegraphics[width=0.48\textwidth]{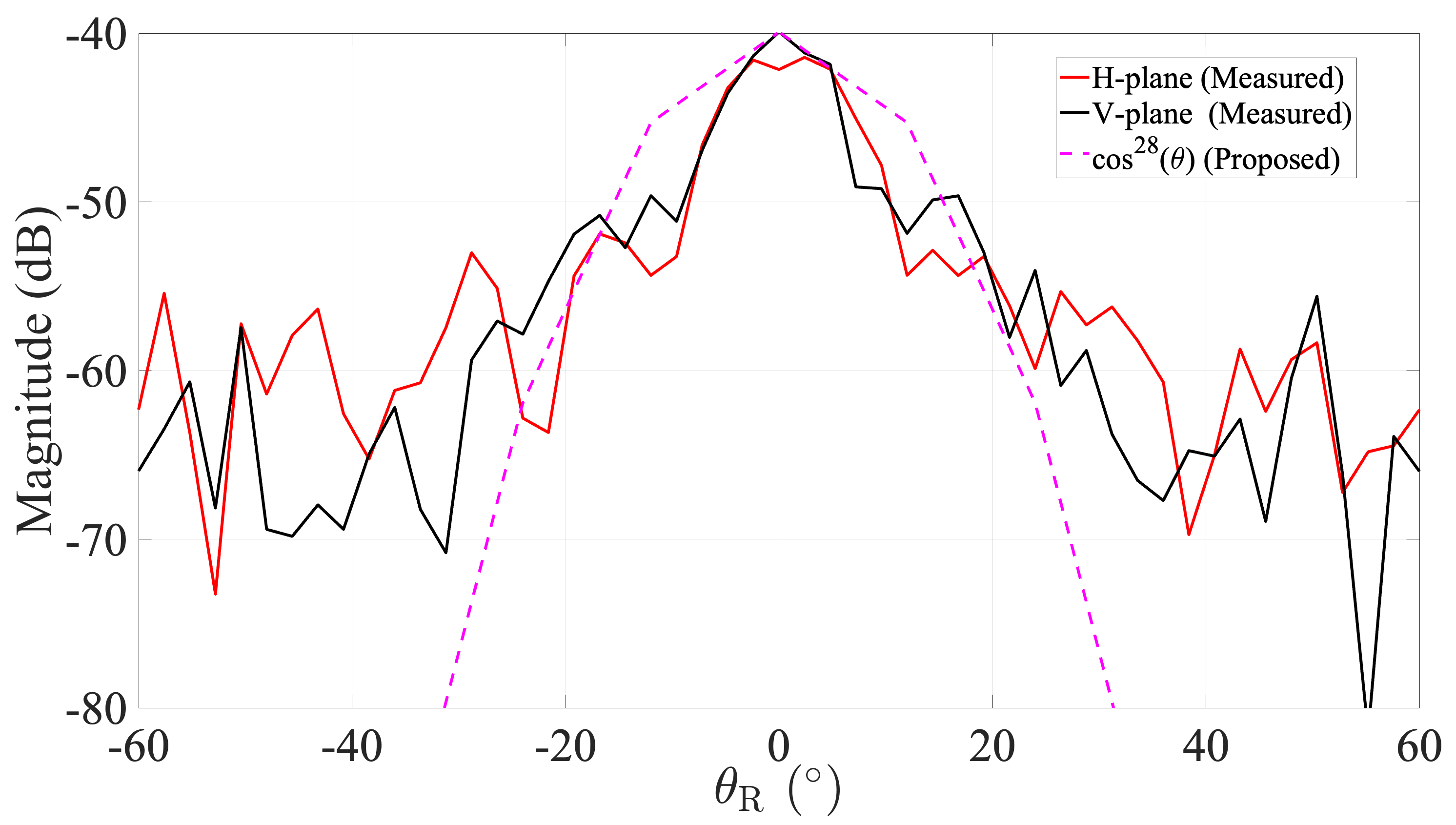}
        \caption{Comparison of magnitude between probe antenna and the cosine-approximation in Stage I: Anechoic chamber verification.}
        \label{fig:magnitude_Measured_cos}
\end{figure}
 We present the comparison between the measured beam pattern and the estimated one with the cosine function in Fig.~\ref{fig:magnitude_Measured_cos}, the normalized cosine-pattern reproduces the measured half-power beamwidth to within a few degrees and aligns closely with the peak boresight amplitude once scaled to the same reference level. $\cos^{28}(\theta)$ is symmetric and free of sidelobes. However, it cannot capture the small amplitude ripples, asymmetries, and null depths observed in measurements. Thus, while the approximation based on cosine function offers an excellent first-order approximation of the main lobe. This is especially useful in theoretical analyses, where more detailed models or measured-data fits are required when sidelobe levels or pattern irregularities must be predicted with high fidelity. 

\textit{Automatically learning to estimate the AoA}: Once we have captured the theoretical approximation of the beam pattern, we can then utilize \ac{sr} to search for the \ac{aoa} with respect to the path loss coefficients. Furthermore, three aforementioned \ac{sr}-based \ac{aoa} estimators are benchmarked, where the performance (in \ac{mae}) and the  final expression, are summarized in Table~\ref {tab:summary_FS}. As we can observe from the results in column~2, Table~\ref{tab:summary_FS}, although the unconstrained \ac{sr} model (row~1) delivers the lowest \ac{mae} ($0.396^\circ$), it does so by combining transcendental and arithmetic operations in a way that offers little physical insight into the underlying propagation or beam pattern. In contrast, SABER explicitly link to known beam patterns (row~2 and~3), with direct inversion having $14$ times better accuracy than the polynomial-based method. Furthermore, there is a trade-off between a minor loss in accuracy and achieving clear interpretability and consistency with theoretical beam pattern behavior. This trade-off indicates that even a minimal amount of prior knowledge, such as the connection between the path loss coefficient and the beam pattern, when integrated with \ac{sr}, results in models that are both highly accurate and easily interpretable according to column~2 and~3 in Table~\ref{tab:summary_FS}. By enforcing the known pattern structure (or by approximating it with a low-order polynomial), we achieve high \ac{aoa} accuracy while maintaining a direct mapping to the antenna design. In contrast, a purely unconstrained \ac{sr}-matching can achieve slightly lower errors of the predicted angles, but at the expense of physical transparency and out-of-sample behavior guarantees.

\subsection{Stage II (RIS-aided systems) results}
\label{subsec:doa}

\textit{Automatically learning to estimate the beam pattern}: We use the same method as in the Stage I to obtain the beam pattern for transmitter and receiver, and we can arrive at $4$ and $1$, respectively. Moreover, with the beamformer, we can assume the \ac{ris} capture all the energy from the transmitter, thus we can assume $\varepsilon_{\mathrm{ap}}=1$. Therefore, we can rewrite the path loss coefficient of RIS-aided system in Eq.~\eqref{eq:PL_RIS_db} as:

\begin{equation}
     \begin{aligned}
      PL_{\mathrm{RIS,tot}}&~[\mathrm{dB}] =10\left(4\log_{10}G_{\mathrm{T,\max}}\cos(\theta_{\mathrm{T}},\phi_{\mathrm{T}})\right.\\
      &\left.+\log_{10}G_{\mathrm{R,\max}}\cos(\theta_{\mathrm{R}},\phi_{\mathrm{R}})\right)\\
        &+20\left(\log_{10}a +\log_{10}b-\log_{10}R_{\mathrm{T,RIS}}\right.\\
        &-\log_{10}R_{\mathrm{RIS,R}}+\left.\log_{10}\cos\theta_{\mathrm{axis}}\right)\\
        &+L_{\mathrm{PD}}+L_{\mathrm{connector}}+L_{\mathrm{cable}}-G_{\mathrm{BF}}.
    \end{aligned}
    \label{eq:PL_RIS_db_final}
\end{equation}

\begin{figure}
    \centering
    \includegraphics[width=0.48\textwidth]{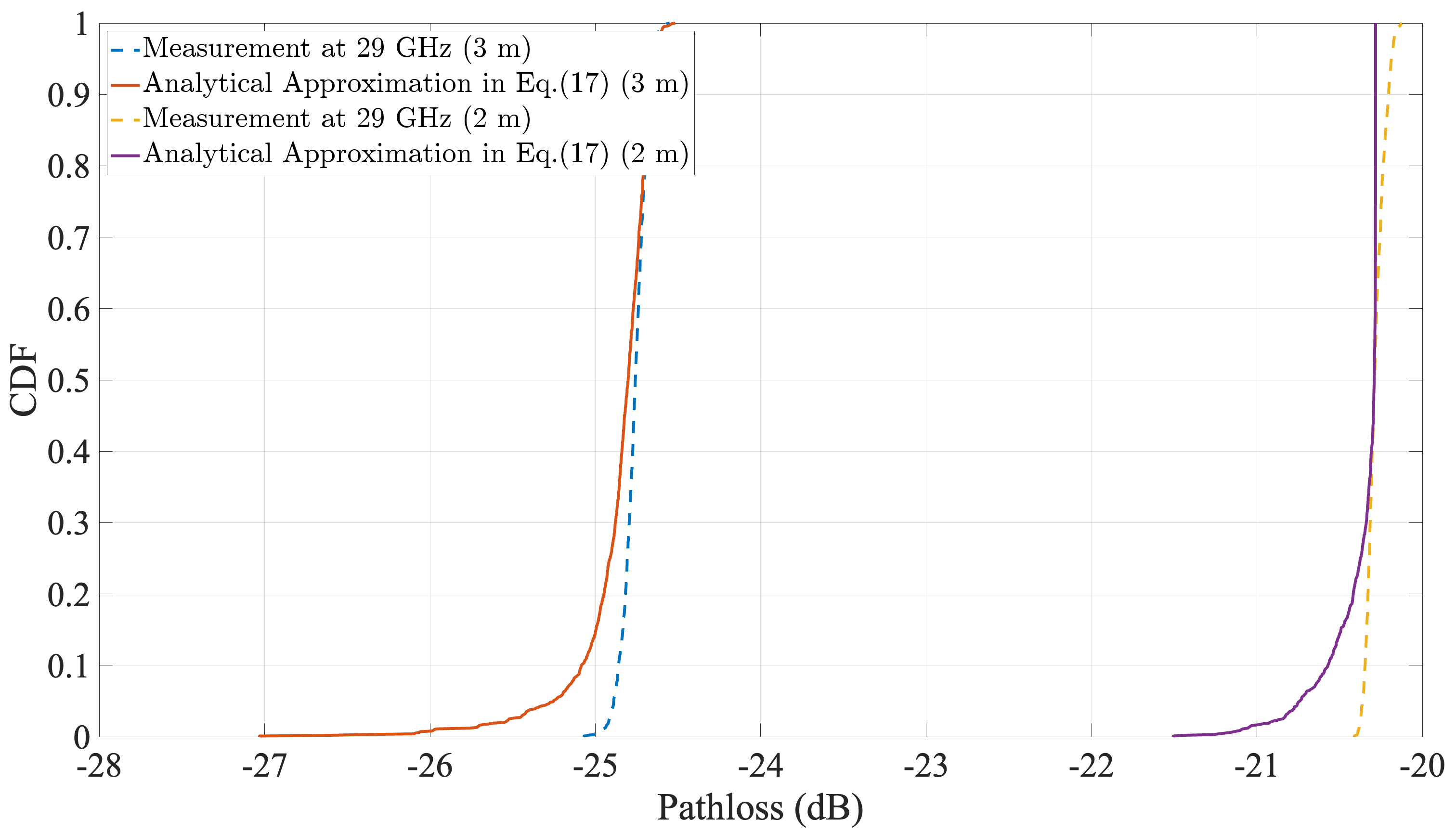}
    \caption{CDF of the measuring data in Stage II and the analytical approximation in Eq.~\eqref{eq:PL_RIS_db_final} via Monte Carlos Simulation.}
    \label{fig:cdf}
\end{figure}

\noindent We take the measurement at 29 GHz as an example to show that the analytical approach in Eq.~\eqref{eq:PL_RIS_db_final} closely reproduces the measured path loss distributions for both $2$~m and $3$~m configurations in Fig~\ref{fig:cdf}. The \ac{cdf}~\cite{CDF_access} slopes and spreads match well, with only a small systematic offset of less than $0.3$~dB for both distances, indicating that the analytical angle-dependent model captures the dominant variability observed in practice. To generate the analytical distributions, a Monte Carlo procedure is applied in which the transmitter and receiver pointing offsets $\theta_{\mathrm{T}}$ and $\theta_{\mathrm{R}}$ are drawn from truncated Gaussian distributions centered at boresight with standard deviation $\sigma_{\mathrm{PL}}=3^\circ$. These angular are then inserted into Eq.~\eqref{eq:PL_RIS_db}, and the process is repeated $3000$ times to obtain a modeled path loss distribution for each distance. 

Since we have already acquired the \ac{aoa} from the setup in Section~\ref{sec:meth_measure}, i.e., $\theta_{\mathrm{R}}=55^{\circ}$. We can take it as ground truth and use \ac{sr}-based method for recovering the information of the angle. Building on the findings from the free-space analysis, the same estimation methodologies are applied to the more complex \ac{ris}-aided propagation environment. The results are shown in the Table.~\ref{tab:summary_RIS}.

\begin{table}[h]
  \centering
  \caption{AoA estimation MAE of three SR-based inversion methods in a RIS-aided indoor setup (identical MAE at 2~m and 3~m for both Tx–RIS and RIS–Rx link).}
  \label{tab:summary_RIS}
  \resizebox{1\columnwidth}{!}{%
    \begin{tabular}{l c c}
      \toprule
      \textbf{Approach} & \textbf{MAE ($^\circ$)} & \textbf{Final expression} ($\hat{\theta}_{\mathrm{R}}$ in radians) \\
      \midrule
      Unconstrained SR                    & $6.53\times10^{-7}$ & 0.9599 \\
      SABER: Direct Inversion             & $6.53\times10^{-7}$ & 0.9599 \\
      SABER: Polynomial-based Cosine Inversion & 0.78 & 0.9736 \\
      \bottomrule
    \end{tabular}%
  }
\end{table}

\textit{Automatically learning to estimate the AoA}: Stage II applies the same \ac{sr}-based inversion methods in the RIS-aided indoor testbed at both 2~m and 3~m. As shown in Table~\ref{tab:summary_RIS}, both the unconstrained \ac{sr} and SABER with Direct Inversion approach (row~1 and~2)converge to the same closed-form cosine model and achieve a virtually zero \ac{mae} ($6.53 \times10^{-7}$) (column~2), effectively recovering the true $55^\circ$ angle with negligible error (column~3). In contrast, the Polynomial-based Cosine Inversion (row~3) fits a low-order surrogate to the cosine response, resulting in a larger error of $0.78^\circ$, which is consistent at both 2~m and 3~m. This confirms that imposing the known beam-pattern structure remains both interpretable and highly accurate even in the more complex, two-hop RIS propagation environment. Meanwhile, the unconstrained \ac{sr} approach in this fixed-angle verification also yielded an interpretable model with near-zero error, suggesting that prior beam-pattern knowledge is not strictly necessary when estimating a single, fixed \ac{aoa}. By contrast, the \ac{aoa} estimation with multi-angle measurement, the same unconstrained \ac{sr} produced highly complex expressions. These results indicate that while unconstrained SR achieves high accuracy in both stages, its ability to balance accuracy and interpretability emerges when estimating a single, deterministic angle.


\subsection{Comparison to \ac{crlb}}
\ac{crlb} is commonly used to benchmark the performance of an estimator, as it establishes the lower bound of \ac{mse} of any unbiased estimator \cite{estimation_theory,CRB_AoA}. This is shown as the last block in Fig.~\ref{fig:flowchart} (highlighted in yellow). For the single-link signal model of Section~\ref{sec:meth_math}, we can then express the received signal as: 

\begin{equation}
    \mathbf{y}=\alpha h(\theta)\mathbf{x}+\mathbf{n},
\end{equation}

\noindent where upper and lower case letters denote matrices and vectors, respectively.  $\theta$ denotes the \ac{aoa}, and $\alpha\!\in\!\mathbb C$ is an angle-independent complex gain capturing all residual amplitude and phase factors not included in $h(\theta)$. More specifically, when a calibration tone at the peak gain direction $\theta_{\mathrm{pk}}$ is available, the magnitude of the unknown gain, $|\alpha|$, is fixed by:

\begin{equation}
    |\alpha| \;=\; \frac{\bigl|S_{21}(\theta_{\mathrm{pk}})\bigr|_{\mathrm{lin}}}{\,h(\theta_{\mathrm{pk}})} ,
    \label{eq:alpha_mag}
\end{equation}

\noindent We assume the phase of $\alpha$ is modelled as uniformly distributed over $[0,2\pi)$. On the other hand, $h(\theta)$ is the deterministic, angle-dependent gain obtained from the path loss expressions in Eq.~\eqref{eq:fspl} and Eq.~\eqref{eq:PL_RIS_db}. Furthermore, $\mathbf{x}$ is the known data stream, with $\|\mathbf{x}\|^{2}=1$. In the absence of a dedicated calibration, we assume the receiver noise ($\mathbf{n}$) follows a zero-mean circularly-symmetric complex Gaussian distribution, i.e., $\mathbf{n}\sim \mathcal{CN}(0,\sigma^{2}\mathbf{I}_{M})$. Since $|\alpha|$ is known, the vector of unknown real-valued parameters is:

\begin{equation}
\boldsymbol{\eta} = [\theta,\phi_{\alpha}, \sigma^2]^T.
\label{eq:eta_vector_correct}
\end{equation}

\noindent The mean received signal is $\mathbb{E}\left\{\mathrm{y}\right\}=\boldsymbol{\mu}(\theta, \phi_\alpha) = |\alpha| e^{j\phi_\alpha} h(\theta) \mathbf{x}$. Its derivatives with respect to the unknown parameters are:

\begin{align}
\frac{\partial \boldsymbol{\mu}}{\partial \theta} &= |\alpha| e^{j\phi_\alpha} h'(\theta) \mathbf{x}, \\
\frac{\partial \boldsymbol{\mu}}{\partial \phi_\alpha} &= j |\alpha| e^{j\phi_\alpha} h(\theta) \mathbf{x},
\end{align}

\noindent where $h'(\theta)=\partial h(\theta)/\partial\theta$. As a result, the complex-Gaussian log-likelihood yields the Fisher information matrix:

\begin{equation}
\begin{aligned}
\mathbf{I}(\theta, \phi_\alpha) 
&= \frac{2M\,|\alpha|^2}{\sigma^2} \\
&\times
\begin{bmatrix}
|h'(\theta)|^2 & -\,\mathrm{Im}\{h'(\theta)^* h(\theta)\} \\
-\,\mathrm{Im}\{h'(\theta)^* h(\theta)\} & |h(\theta)|^2
\end{bmatrix}.
\end{aligned}
\label{eq:FIM_single_link}
\end{equation}

\noindent where $\mathrm{Im}\left\{\cdot\right\}$ represent taking the imaginary part of the variable. By inverting $\mathbf{I}(\theta, \phi_\alpha)$ yields the general \ac{crlb} for $\theta$:
\begin{equation}
\mathrm{Var}\{\hat{\theta}\} \;\ge\; 
\frac{\sigma^2}{2M\,|\alpha|^2}\,
\frac{|h(\theta)|^2}{|h(\theta)|^2\,|h'(\theta)|^2 - \left(\mathrm{Im}\{h'(\theta)^* h(\theta)\}\right)^2}.
  \label{eq:CRLB_theta}
\end{equation}

When the beam-pattern phase is constant with respect to $\theta$. i.e., $\mathrm{Im}\{h'(\theta)^* h(\theta)\} = 0$. Eq.~\eqref{eq:CRLB_theta} can be reduced to the simpler form:
\begin{equation}
\mathrm{Var}\{\hat{\theta}\} \;\ge\;
\frac{\sigma^2}{2M\,|\alpha|^2\,|h'(\theta)|^2}.
\label{eq:CRLB_simplified}
\end{equation}

Taking the square root gives the RMSE form:
\begin{equation}
\sqrt{\mathrm{CRLB}(\theta)} = \sqrt{\frac{\sigma^2}{2M\,|\alpha|^2\,|h'(\theta)|^2}}\,.
\label{eq:CRLB_RMSE}
\end{equation}





\begin{figure}
    \centering
    \includegraphics[width=0.48\textwidth]{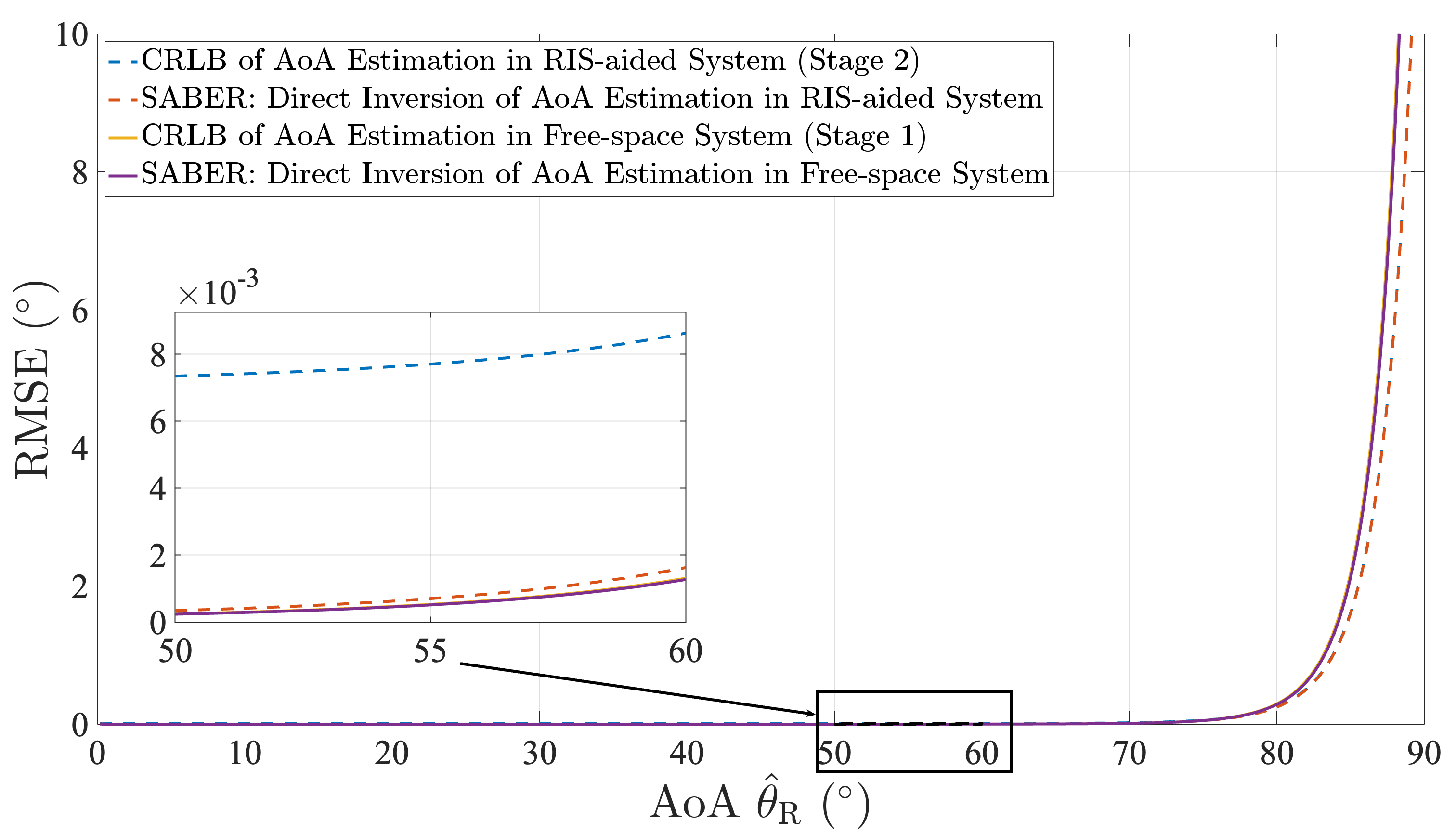}
    \caption{Empirical RMSE of SABER: Direct Inversion compared with the CRLBs of each stage, where free-space is shown in solid lines, while RIS-aided case is shown in dash lines.}
    \label{fig:CRLB}
\end{figure}

\noindent Finally, by inserting the explicit form of the beam‐pattern gain \(h(\theta)\) and its derivative \(h^{\prime}(\theta)\) given by our closed‐form path loss expressions in Eqs.~\eqref{eq:PL_RIS_db} and ~\eqref{eq:path_loss_dB_theta_phi_SR}, we obtain a completely analytical curve $\sqrt{\mathrm{CRLB}(\theta)}$ that depends only on known system parameters, the measurement noise variance $\sigma^2$, and the number of snapshots $M$. The result of comparing \ac{crlb} to the actual \ac{rmse} of the estimation is shown in Fig.~\ref{fig:CRLB}, we assume the noise variance \(\sigma^{2}=10^{-3}\) and the number of snapshots $M$ equal to $1000$. The results of free-space (Stage~I) and \ac{ris}-aided system (Stage~II) are shown in solid and dashed lines, respectively. Moreover, the inset highlights the region between  $\hat{\theta}_{R}=50^\circ$ and $\hat{\theta}_{R}=60^\circ$, where the difference between the \ac{crlb} and the achieved \ac{rmse} is on the order of $10^{-3}$ degrees. In particular, SABER: Direct Inversion in Stage I is almost indistinguishable from the \ac{crlb} curve, while in Stage II, a slight but consistent gap remains, with SABER (red dashed line) following the same trend as the \ac{ris}-aided \ac{crlb} but at a marginally higher \ac{rmse} level. This close agreement demonstrates that SABER operates near the fundamental estimation limit for both tested configurations, confirming their statistical efficiency and robustness even under different propagation conditions. As expected from the derivative term in the \ac{crlb} expression, the bound sharply increases as $\theta$ approaches $90^\circ$, reflecting the loss of angular sensitivity when the array is illuminated from the direction that is parallel to the radiation pattern of the received terminal. This behavior is not solely a theoretical phenomenon; from a hardware perspective, handling large-angle oblique incidence remains a significant challenge. Even in advanced beam-steering systems, achieving reliable performance at extreme steering angles is difficult, highlighting the practical relevance of the divergence near $90^\circ$~\cite{xie2016oam}, which can cause beam alignment ill-conditioned and increases the risk of mispointing.

\section{Conclusion and Future Work}
\label{sec:conclusion}

In this work, we introduced SABER, a \ac{sr}-based solution for \ac{aoa} estimation that relies solely on the measured path loss coefficient. Specifically, we employed a \ac{sr}-based \ac{ml} framework to recover antenna directivity and beam-pattern characteristics, enabling interpretable \ac{aoa} estimation in both a controlled free-space setup and a realistic \ac{ris}-aided indoor testbed. Unlike many \ac{ml}-based methods, SABER achieves not only high accuracy but also interpretable closed-form models, facilitating future integration into physics-aware design and analysis. More specifically, in Stage I, covering a multi-angle sweep in free space, we showed that embedding minimal prior knowledge: either the exact $\cos^n$ form or a low-order polynomial surrogate, as well as, the relationship between path loss coefficient and the \ac{aoa}. We can preserve near-optimal performance ($\mathrm{MAE}=0.42^\circ$), while guaranteeing physical transparency and consistent extrapolation. In contrast, unconstrained \ac{sr} achieved the lowest prediction error ($\mathrm{MAE}=0.396^\circ$) but produced highly complex, non-intuitive expressions. On the other hand, in Stage II, where the \ac{aoa} was fixed and deterministic in a \ac{ris}-aided setup, the unconstrained \ac{sr} approach achieved both high accuracy ($\mathrm{MAE}=6.53\times 10^{-7}$) and interpretability, indicating that prior beam-pattern knowledge is not strictly necessary in such single-angle estimation tasks. These findings demonstrate that unconstrained SR maintains high accuracy across both scenarios, but its interpretability depends on the nature of the estimation problem: prior knowledge is valuable for multi-angle estimation, whereas simpler, fixed-angle cases can achieve a natural balance between accuracy and interpretability without it. Moreover, by benchmarking against the theoretical \ac{crlb} bounds, SABER showed near-optimal performance. 

As for future work, we will consider more advanced deep \ac{sr} methods, such as, neural‐guided expression search to further automate model discovery and compare their complexities. Moreover, validate the resilience of the \ac{sr}-based framework in more challenging propagation scenarios, including \ac{mimo} indoor channels and outdoor multi-path environments, where richer channel statistics and temporal dynamics must be accommodated.


\bibliography{bibliography}
\end{document}